\pdfoutput=1

\documentclass[11pt,twoside,a4paper,cmspaper,final,collab]{cms-tdr}
\def\svnVersion{115740:115742}\def\cmsCernNo{2012-104}\def\cmsCernDate{\today}\def\cmsMessage{Submitted to the Journal of High Energy Physics}

\begin{document}\cmsNoteHeader{TOP-11-019}

\hyphenation{had-ron-i-za-tion}
\hyphenation{cal-or-i-me-ter}
\hyphenation{de-vices}

\RCS$Revision: 115742 $
\RCS$HeadURL: svn+ssh://alverson@svn.cern.ch/reps/tdr2/papers/TOP-11-019/trunk/TOP-11-019.tex $
\RCS$Id: TOP-11-019.tex 115742 2012-04-12 18:32:28Z alverson $
\cmsNoteHeader{TOP-11-019} 
\title{Measurement of the mass difference between top and antitop quarks}

\date{\today}

\abstract{A measurement of the mass difference between the top and the antitop quark ($\Delta m_{\text{t}} = m_{\mathrm{t}} - m_{\overline{\mathrm{t}}}$) is performed using events with a muon or an electron and at least four jets in the final state. The analysis is based on data collected by the CMS experiment at the LHC, corresponding to an integrated luminosity of $4.96 \pm 0.11$\fbinv, and yields the value of $\Delta m_{\mathrm{t}} = -0.44 \pm 0.46 ~\stat \pm 0.27 ~\syst\GeV$. This result is consistent with equality of particle and antiparticle masses required by CPT invariance, and provides a significantly improved precision relative to existing measurements.}

\hypersetup{%
pdfauthor={CMS Collaboration},%
pdftitle={Measurement of the mass difference between top and antitop quarks},%
pdfsubject={CMS},%
pdfkeywords={CMS, physics}}

\maketitle 

\section{Introduction and overview} \label{sec:intro}

The standard model of particle physics is a local gauge-invariant quantum field theory in which symmetries play a fundamental role that includes the dependence of system properties under specific transformations such as charge conjugation (C), parity or space reflection (P) and time reversal (T). These individual symmetries and the combined CP symmetry are known to be violated in weak interactions, but the CPT combination appears to be conserved in nature~\cite{PDG}. A major consequence of CPT conservation is that the mass of any particle must equal that of its antiparticle. We focus on a measurement of the mass difference between the top and antitop quark. Since quarks carry color charge and hadronize into colorless particles before decaying, they cannot be observed as free quarks. The lone exception is the top quark, which due to its short lifetime decays before hadronization. 
The mass difference between the top quark and its antiquark
was measured previously by the \DZERO and CDF experiments, and showed no significant deviation from zero~\cite{D0massDiffOrg,CDFmassDiff,D0massDiff}.

This letter reports a measurement of the difference between the mass of the top quark (t) and of its antiparticle ($\overline{\text{t}}$), with significantly reduced uncertainties, using \ttbar events produced in proton-proton collisions at $\sqrt{s} = 7$ TeV, recorded with the Compact Muon Solenoid (CMS) detector at the Large Hadron Collider (LHC)~\cite{lhc}. We select events where one W boson, either from the top or antitop quark, decays into $\text{q} \overline{\text{q}}'$ ($\text{t} \to \text{bW}^+ \to \text{bq} \overline{\text{q}}'$, or its charge conjugate), and the other W decays leptonically ($\text{t} \to \text{bW}^+ \to \text{b}\ell^+\nu_{\ell}$, or its charge conjugate), where the lepton $\ell$ is a muon or an electron. The data are split into $\ell^-$ and $\ell^+$ samples that contain, respectively, three-jet decays of the associated top or antitop quarks. For each event category, the Ideogram likelihood method~\cite{Abdallah:2008xh} is used to measure the mass of the top quark ($m_{\rm t}$) or antitop quark ($m_{\overline{\rm t}}$), and the difference between the masses in the two categories of lepton charge is taken as the mass difference $\Delta m_{\text{t}} \equiv m_{\text{t}} - m_{\overline{\text{t}}}$. The Ideogram method was used previously~\cite{Abazov:2007rk,Aaltonen:2006xc} to measure the mass of the top quark. The procedure incorporates a kinematic fit of the events to a \ttbar hypothesis that is modified specifically for this analysis to consider only the top or antitop quark that decays to three jets.

\section{The CMS detector} \label{sec:CMS}

The central feature of the CMS apparatus is a superconducting solenoid of 6~m internal diameter, providing a magnetic field of 3.8~T. The field volume houses the silicon-pixel and silicon-strip trackers, a crystal electromagnetic calorimeter (ECAL) and a brass/scintillator hadron calorimeter. The inner tracker reconstructs charged-particle trajectories within the pseudorapidity
range $|\eta| < 2.5$, where the pseudorapidity is defined in terms of the polar angle $\theta$ relative to the counterclockwise-rotating proton beam as $\eta \equiv - \ln{ ( \tan{ \theta / 2 } ) }$. The tracker provides an impact parameter resolution of $\approx 15~\mu$m and a transverse momentum ($p_{\rm T}$) resolution of $\approx 1.5\%$ for 100\GeV particles.
The energy resolution is $< 3\%$ for the electron energies in this analysis. Muons are measured for $|\eta|< 2.4$ using gaseous detection planes based on three technologies: drift tubes, cathode-strip and resistive-plate chambers. Matching outer muon trajectories to tracks measured in the silicon tracker provides a transverse momentum resolution of $1 - 6\%$ for the $p_{\rm T}$ values in this analysis. In addition to the barrel and endcap detectors, CMS has extensive forward calorimetry. A more detailed description of the CMS detector can be found in Ref.~\cite{CMS:2008zzk}.

\section{Data and simulation} \label{sec:Data}

This analysis is based on a data sample corresponding to an integrated luminosity of $4.96 \pm 0.11$~\fbinv collected in \Pp\Pp~collisions at a center-of-mass energy of 7\TeV and recorded with the CMS detector. Events are selected through a trigger requiring an isolated electron or muon with $\PT>25$ or $17$\GeV, respectively, accompanied by at least three jets of $\PT>30$\GeV in each event. The acquired data are compared to a set of simulated \Pp\Pp~collisions at $\sqrt{s} = 7$\TeV. Most signal and background events are generated with the matrix-element generator \MADGRAPH 4.4.12~\cite{MadGraph}, interfaced to \PYTHIA 6.4.22~\cite{pythia} for the parton showering, where \ttbar events are generated accompanied by up to three extra partons. The \textsc{mlm} algorithm~\cite{MLMmatching} is used for matching the matrix-element partons to their parton showers. Singly produced top-quark events are generated with the \textsc{powheg} event generator~\cite{Powheg} and generic multijet events with \PYTHIA. The simulation of multijet events is used just to normalize a multijet-enriched control sample of data needed in the analysis (described below). The simulation also includes effects of pileup in \Pp\Pp~collisions, which refers to additional \Pp\Pp~interactions that can occur during the same bunch crossing or in those immediately preceding or following the primary generated process. The simulated event samples are normalized to the theoretical cross section for each process, as calculated with \textsc{fewz}~\cite{FEWZ} for W and Z production, with \PYTHIA for multijet production, and \MCFM~\cite{MCFM} for all other contributing processes. The generated events are then passed through the full CMS detector simulation based on \textsc{geant4}~\cite{Geant4}, and eventually reconstructed using the same algorithms as used for data.

\section{Event reconstruction and selection} \label{sec:selection}

Events are reconstructed using the particle-flow (PF) algorithm~\cite{CMS-PAS-PFT-10-002}, which combines the information from all CMS sub-detectors to identify and reconstruct individual particles produced in the proton-proton collision. The reconstructed particles include muons, electrons, photons, charged and neutral hadrons. Muons are reconstructed using the combined information from the silicon tracker and muon system~\cite{CMS-PAS-MUO-10-002}. Electron reconstruction starts from energy depositions in the ECAL, which are then matched to hits in the silicon tracker and used to initiate a track reconstruction algorithm. This algorithm takes into account the possibility of significant energy loss of the electron through bremsstrahlung as it traverses the material of the tracker~\cite{CMS-PAS-EGM-10-004}. Charged particles are required to originate from the primary collision vertex, identified as the reconstructed vertex with the largest value of $\Sigma p_{\rm T}^2$ for its associated tracks. The list of charged and neutral PF particles originating from the primary collision vertex is used as input for jet clustering based on the anti-\kt algorithm~\cite{antikt} with a distance parameter of $0.5$. Particles identified as isolated muons and electrons are excluded from jet clustering. The momentum of a jet is determined from the vector sum of all particle momenta in the jet, and from simulation found to be typically within $5 - 10\%$ of the true jet momentum. Jet-energy-scale corrections are applied to all the jets in data and simulation. Jets in data have a residual correction that is determined from an assumed momentum balance in dijet and photon+jet events. These corrections are defined as a function of \PT and $\eta$ of the reconstructed jet so as to obtain a more uniform energy response at the particle level, which tends to equalize the jet response in data and simulation~\cite{JESpaper}. The energy of jets is also corrected for the presence of additional pileup from neutrals, as the neutral component of pileup is still present after rejecting the contribution from charged hadrons.

Events in the \Pgm+jets channel are required to contain only one isolated muon with $\PT>20$\GeV and $|\eta| < 2.1$, while the e+jets channel requires only one isolated electron with $\PT>30$\GeV and $|\eta|<2.5$. The relative isolation $I_{\rm rel}$ is calculated from the other PF particles within a cone of $\DR = \sqrt{(\Delta \eta)^2 + (\Delta \phi)^2} < 0.4$ around the axis of the lepton, with $\phi$ representing the azimuthal angle. It is defined as $I_{\rm rel} = (I_{\rm charged} + I_{\rm photon} + I_{\rm neutral}) / \PT$, where $I_{\rm charged}$ is the transverse energy deposited by charged hadrons, and $I_{\rm photon}$ and $I_{\rm neutral}$ are the respective transverse energies of photons and neutral particles not identified as photons. Leptons are considered to be isolated when $I_{\rm rel} < 0.125$. Furthermore, events must have at least four reconstructed jets with $\PT > 30$\GeV and $|\eta| < 2.4$. Additional event selection criteria are discussed in Section~\ref{sec:kinfit}. Table~\ref{table:eventCountsSplitted} gives the number of events observed in data following all selections, and the number expected from simulation, separately for events with $\mu^+$, $\mu^-$, e$^+$ and e$^-$. The kinematic characteristics of the multijet background are estimated from a data sample of events that pass all selections, but with an inverted lepton-isolation criterion of $I_{\rm rel} > 0.2$. The number of expected multijet events passing all selections is normalized to the \PYTHIA simulation.
\begin{table}[htb]
  \begin{center}
    \caption{Number of events following full selection of \Pgmp+jets, \Pgmm+jets, \Pep+jets and \Pem+jets events in data, and the expectation from simulations before any rescaling. Uncertainties are purely statistical and do not include contributions from production cross sections, integrated luminosity, detector acceptance, or selection efficiencies, as discussed in Refs.~\cite{CMS-PAS-TOP-10-002,CMS-PAS-TOP-10-003}. \label{table:eventCountsSplitted}}
    \begin{tabular}{lcccc}
      \hline \hline
      Sample               &   \Pgmp+jets &  \Pgmm+jets &  \Pep+jets &  \Pem+jets \\ \hline
      \ttbar               & 15028 $\pm$ 56  & 15006 $\pm$ 56  & 10649 $\pm$ 47  &  10611 $\pm$ 47  \\
      W+jets               & 11180 $\pm$ 149 &  7314 $\pm$ 121 &  7783 $\pm$ 125 &  5523 $\pm$ 105  \\
      Z$/ \gamma^{*}$+jets &  1410 $\pm$ 25  &  1516 $\pm$ 26  &  1607 $\pm$ 27  &  1685 $\pm$ 27  \\
      Single top           &   951 $\pm$ 7   &   850 $\pm$ 7   &   675 $\pm$ 6   &   610 $\pm$ 6   \\
      Multijet             &   483 $\pm$ 90  &   196 $\pm$ 57  &  722 $\pm$ 246  &   1413 $\pm$ 485  \\ 
      Total                & 29050 $\pm$ 185 & 24882 $\pm$ 147 & 21436 $\pm$ 281 & 19842 $\pm$ 499 \\ \hline 
      Observed             & 27038 & 23928 & 22999 & 21111 \\ \hline \hline
    \end{tabular}
  \end{center}
\end{table}

\begin{figure*}[!htb]
  \centering
  \subfloat[]{\includegraphics[width=0.45 \textwidth]{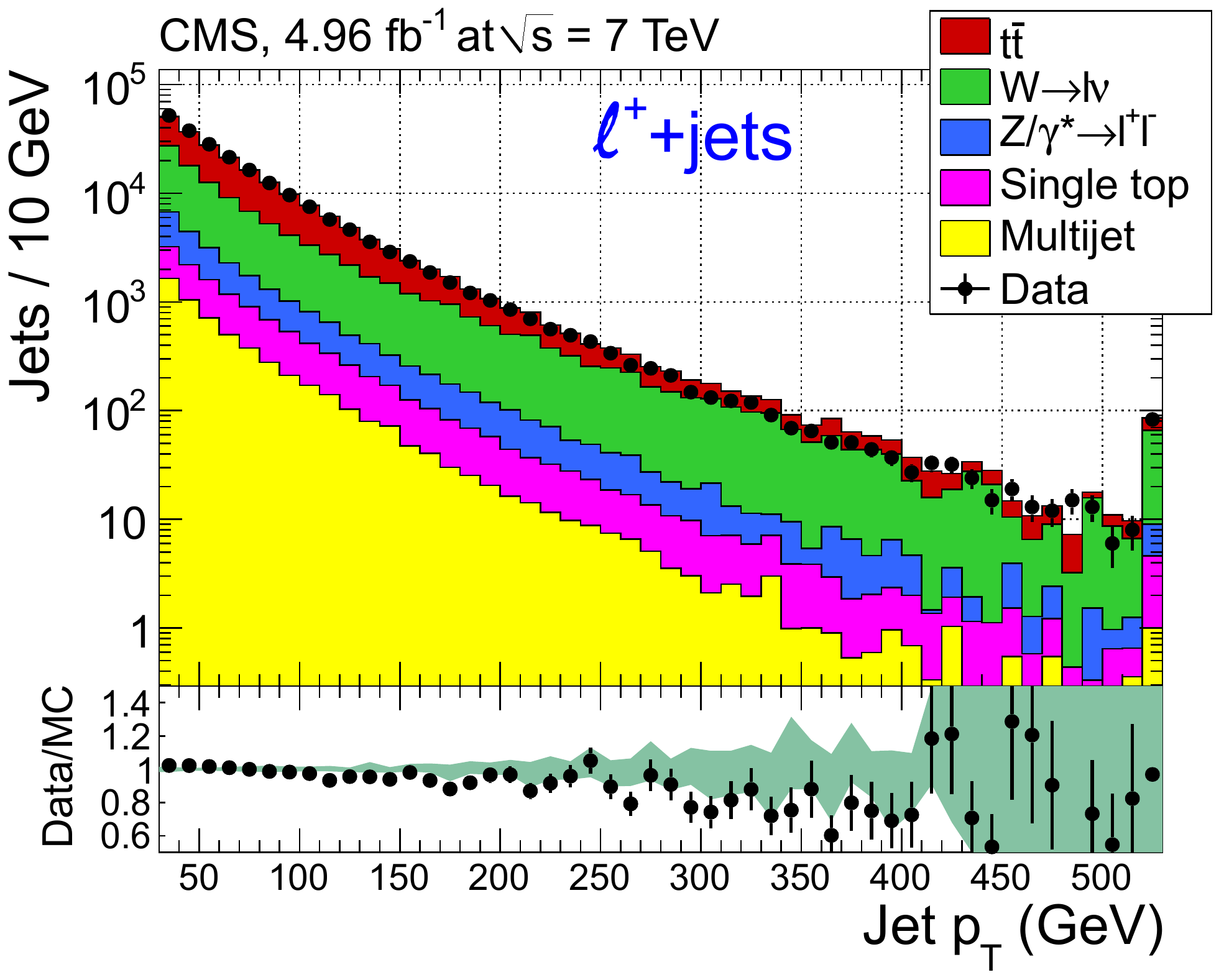} \label{fig:dataMCselection1}}
  \subfloat[]{\includegraphics[width=0.45 \textwidth]{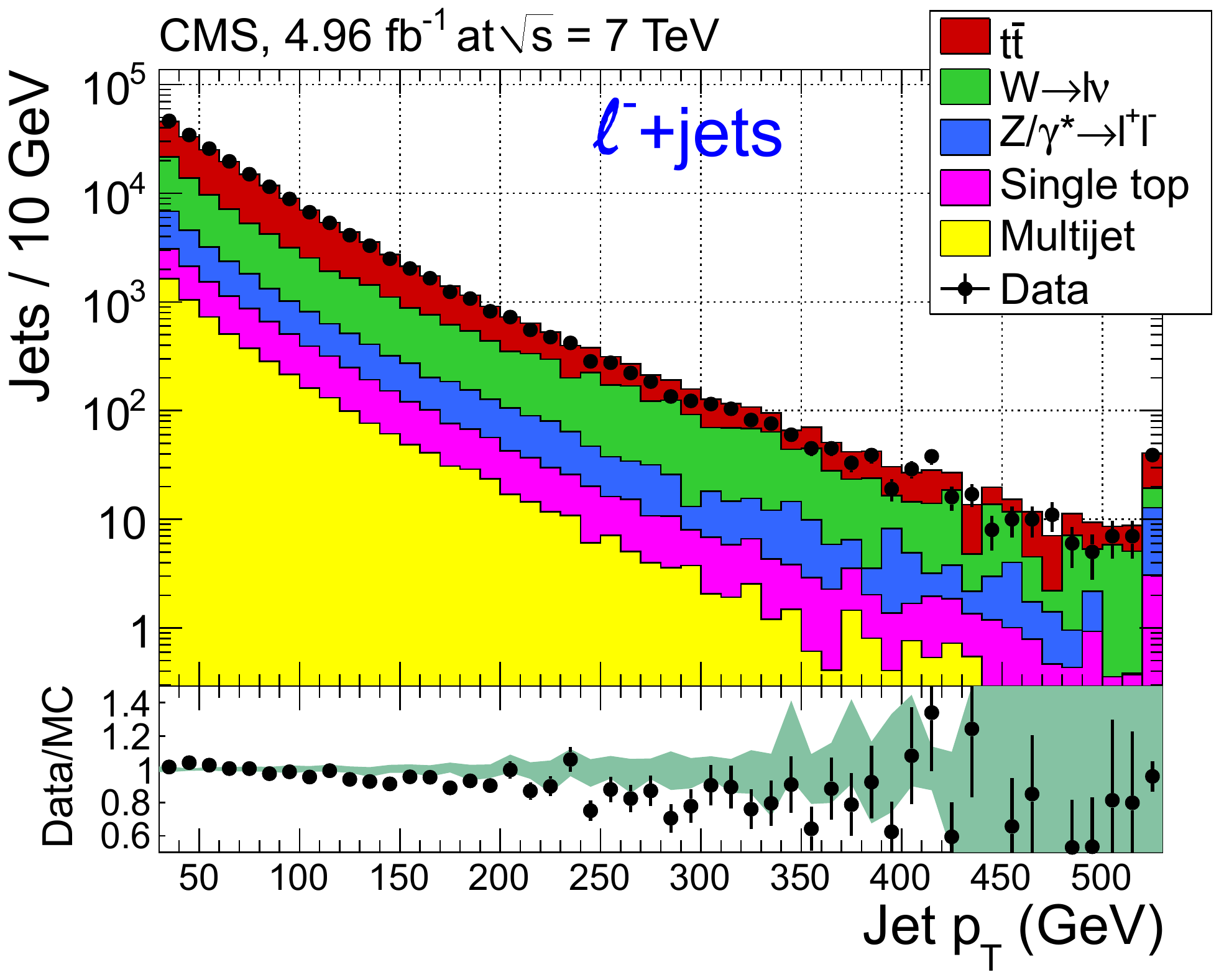} \label{fig:dataMCselection2}} \\ \vspace{0.3cm}
  \subfloat[]{\includegraphics[width=0.45 \textwidth]{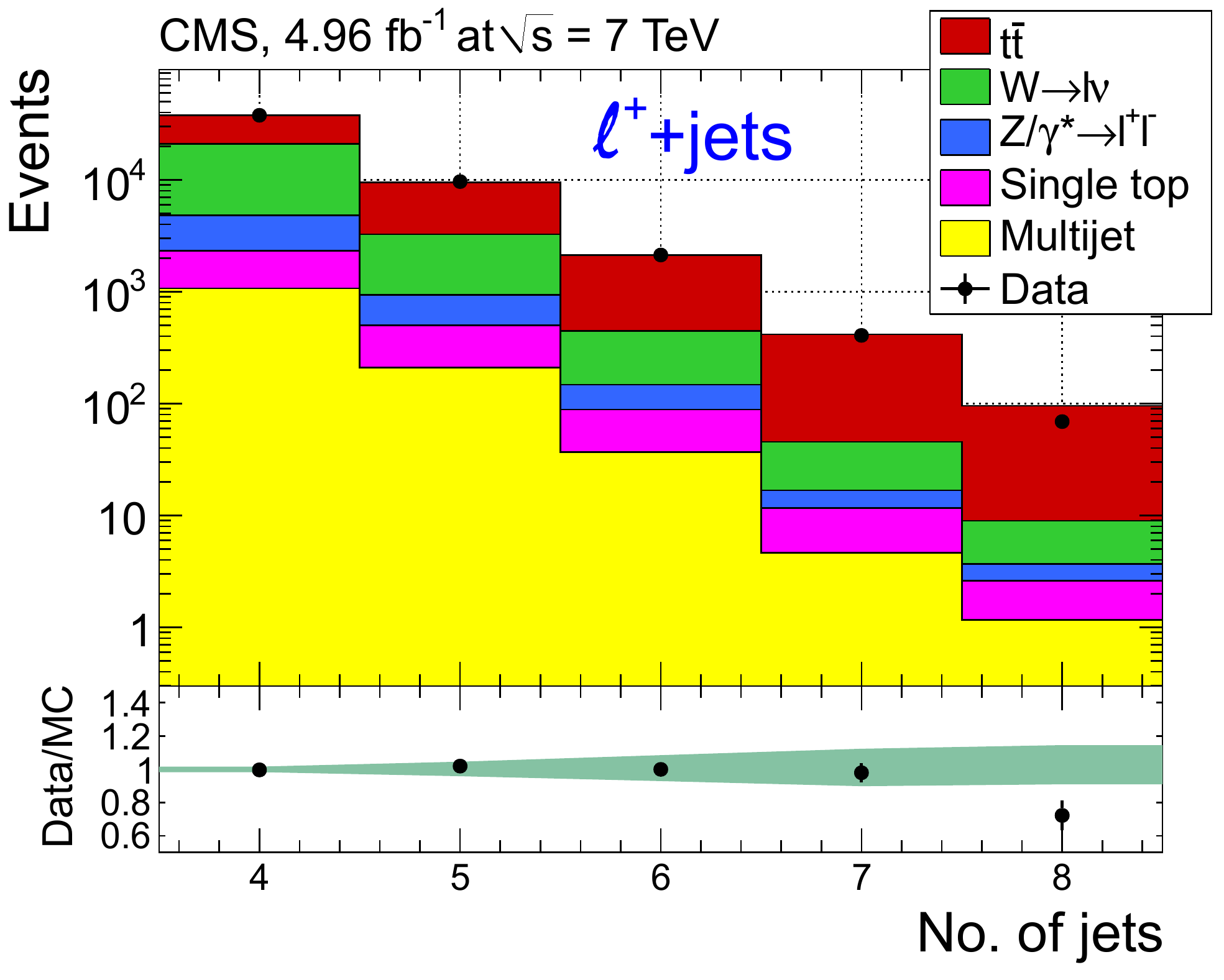} \label{fig:dataMCselection3}}
  \subfloat[]{\includegraphics[width=0.45 \textwidth]{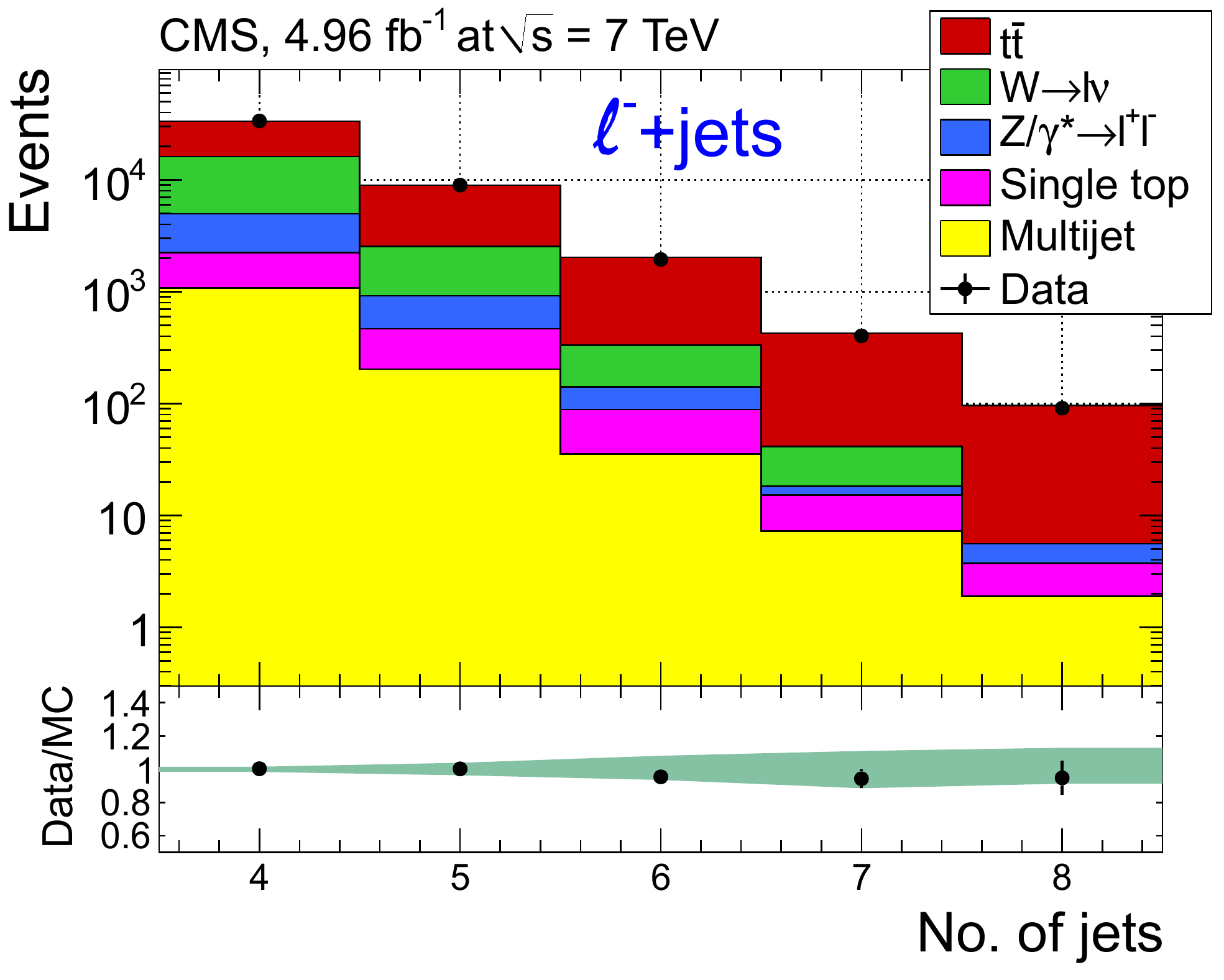} \label{fig:dataMCselection4}}
  \caption{
  Panels \protect\subref{fig:dataMCselection1} and \protect\subref{fig:dataMCselection2} display the transverse momenta of all jets in \ttbar events, for $\ell^+$ and $\ell^-$ events, respectively, while panels \protect\subref{fig:dataMCselection3} and \protect\subref{fig:dataMCselection4} give the number of selected jets per event that pass all selections. The simulation is normalized to the number of events observed in data. Overflows are included in the last bins of the distributions. The ratio of the number of observed events in data to the number of events expected from simulation is shown at the bottom of each plot. The error band corresponds to the systematic uncertainties related to jet energy scale, jet energy resolution, background estimation and modeling of pileup.
  \label{fig:dataMCselection}
  }
\end{figure*}

Agreement between data and simulation in the number of selected events (normalization) is less important for this analysis than agreement for their kinematic distributions. The simulated signal and background events are therefore rescaled through a single global factor to match the number of events observed in data, keeping the relative background fractions fixed to the expectations from simulation. After this rescaling, a comparison of simulation and data for several key distributions is shown in Fig.~\ref{fig:dataMCselection}. In general, the data appear to be well modeled by the simulation. The small possible deviations between data and simulation at large jet \PT values have little impact on this analysis, as most \ttbar events have jet transverse momenta below 200\GeV.

\section{Kinematic fit} \label{sec:kinfit}

A kinematic fit of $\ell$+jets final states to a \ttbar hypothesis, applying the constraints of transverse-momentum conservation, the assumed equality of $m_{\text{t}}$ and $m_{\overline{\text{t}}}$, and the accepted value of 80.4\GeV for the mass of the W boson ($m_{\text{W}}$), has been one of the successful methods for extracting the mass of the top quark from \ttbar events. The basic features of this type of kinematic fit are described in Refs.~\cite{ThesisPetra,KinFitNote}. The fit we use corresponds to a modification that reconstructs the mass of the three-jet decays of top quarks ($\text{t} \to \text{bW} \to \text{bq} \overline{\text{q}}'$) by varying the momenta of the two jets that are assigned to the W boson, using $m_{\text{W}}$ as a constraint. For each event, the four jets with highest transverse momentum (leading jets) are considered in the fit.
These four jets can be associated with the four quarks for the \ttbar-decay hypothesis $\ttbar \rightarrow \text{b}\overline{\text{b}}\PWp\PWm \rightarrow \text{b}\overline{\text{b}} \text{q} \overline{\text{q}}' \ell\nu_{\ell}$ in 24 possible ways. However, since the interchange of the two quarks from W-boson decay ($\text{q} \overline{\text{q}}'$) offers the same mass information, only 12 of these combinations provide unique solutions. The four leading jets do not always originate from the quarks of the \ttbar decay, because of the presence of additional jets from gluon radiation. In simulated \ttbar events, the three quarks arising from the top quark that decays to three jets are among the four leading jets in $\approx 70$\% of all such $\ell$+jets events.

The kinematic fit is performed for each of the 12 jet-to-quark assignments. However, before implementing the fit, additional corrections are applied to correct jet energies to the parton level. These are derived separately for light-quark jets and for b jets in bins of $|\eta^{\rm jet}|$ and $\PT^{\rm jet}$, by comparing the transverse energies \et of selected jets with the \et of generated partons in simulated \ttbar events. The correction factors depend on the flavor of each jet for a given jet-to-quark assignment, and are about 4\% larger for b jets than for light-quark jets. The parton-corrected jets used as input for the kinematic fit are parametrized by their $\et$, $\theta$, and $\phi$. The resolutions of the reconstructed jet quantities are also used as input, and are obtained from the width of the distributions for differences in \et, $\theta$, and $\phi$ between parton-corrected jets and the nascent parton values. As indicated above, the kinematic fit adjusts the momenta of the two light jets, taking their corresponding resolutions into account, while keeping the $E/p$ of each jet fixed.
Only solutions with $\chi^2/n_{\rm dof} < 10$ are accepted, where $n_{\rm dof}$ ($= 1$) corresponds to the number of degrees of freedom in the fit. An event is rejected if no combination of jets passes the $\chi^2/n_{\rm dof}$ requirement. The efficiency of this requirement in simulated \ttbar events is 88\%. The most important gain from the kinematic fit is that it improves the resolution on the top-quark mass. For correct jet combinations, the mass resolution is improved from $\approx$~15\GeV to $\approx$~10\GeV, as estimated from simulated \ttbar events with $m_t = 172.5$\GeV.

\begin{figure*}[!htb]
  \centering
  \subfloat[]{\includegraphics[width=0.45 \textwidth]{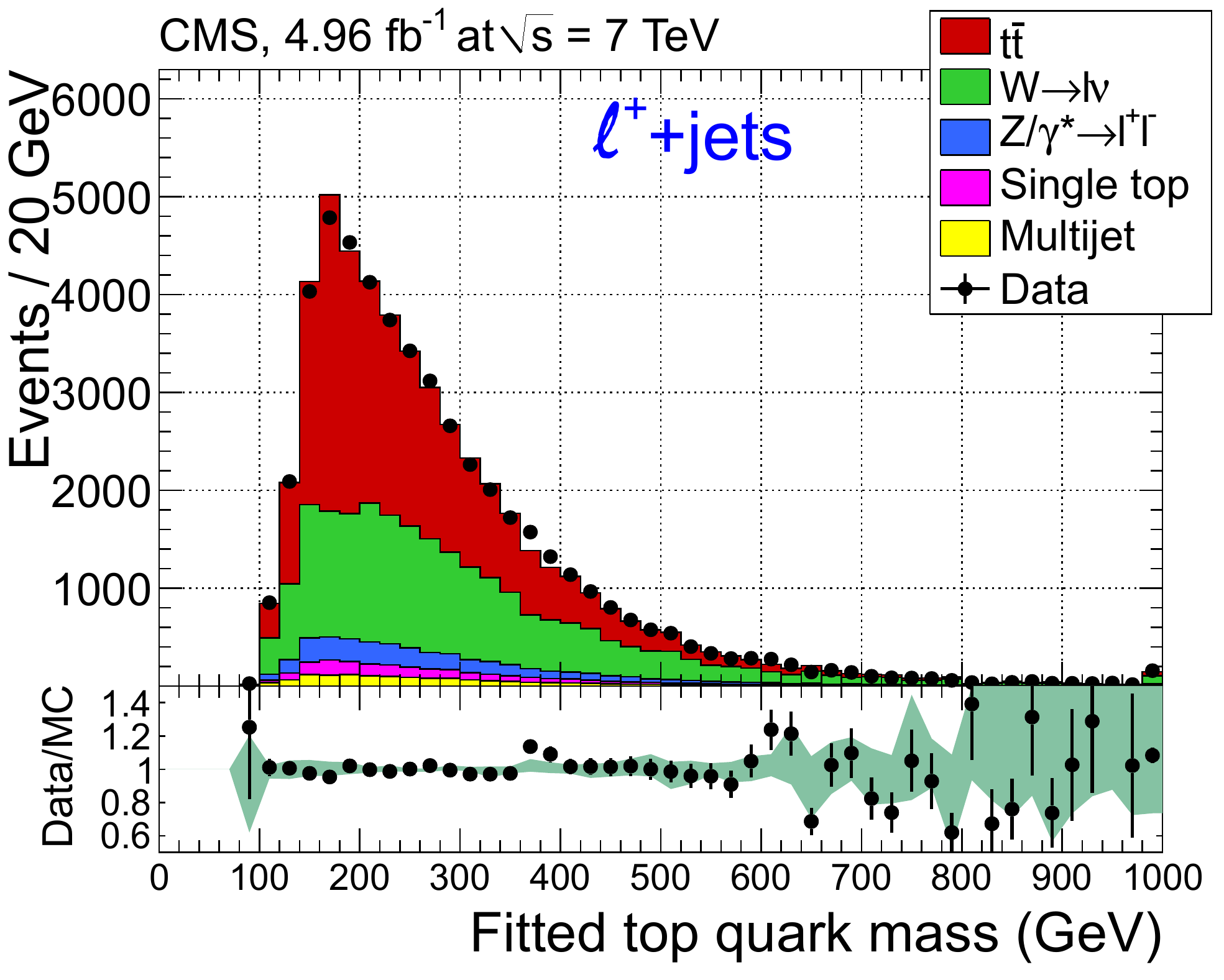} \label{fig:dataMCkinfit1}}
  \subfloat[]{\includegraphics[width=0.45 \textwidth]{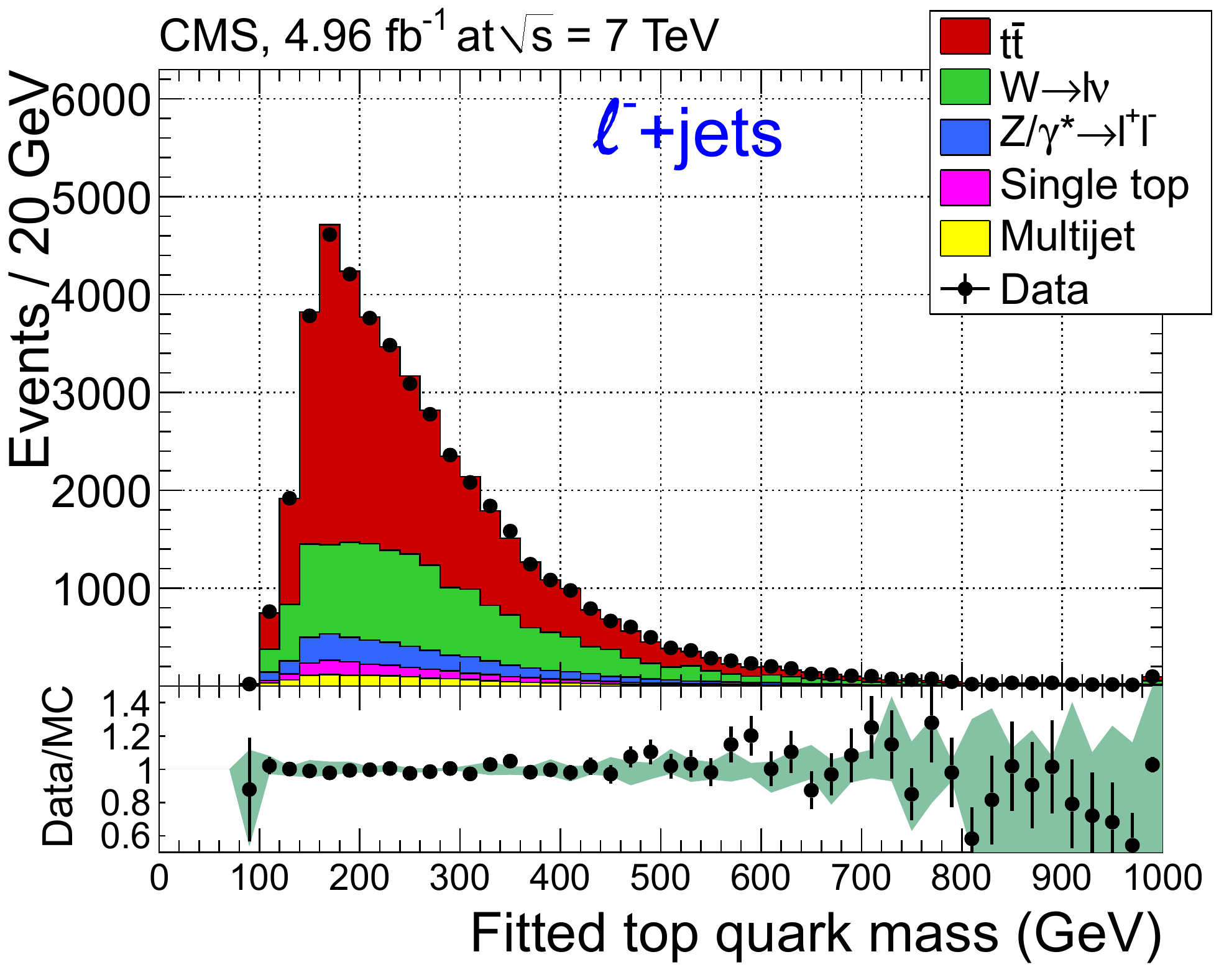} \label{fig:dataMCkinfit2}} \\ \vspace{0.3cm}
  \subfloat[]{\includegraphics[width=0.45 \textwidth]{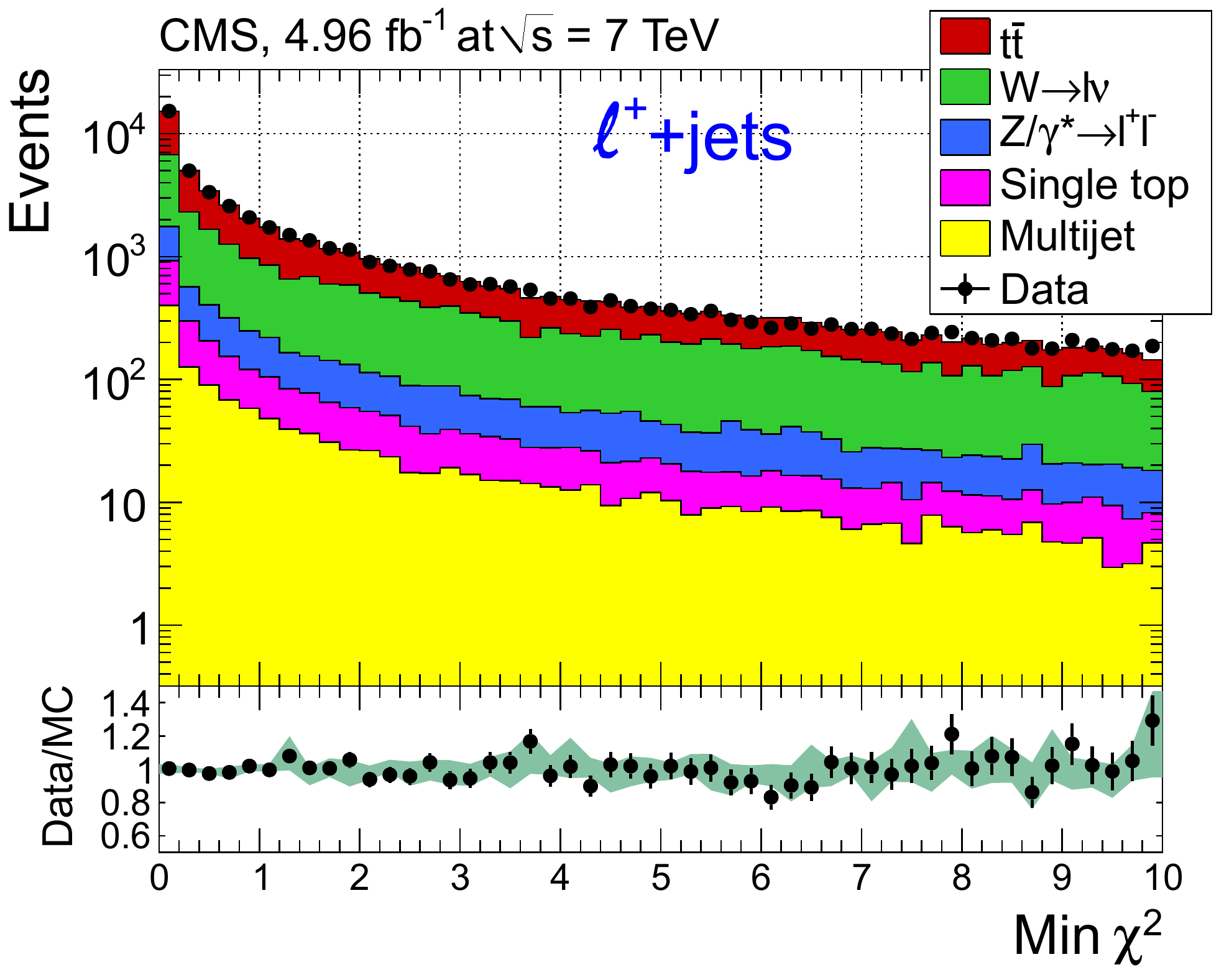} \label{fig:dataMCkinfit3}}
  \subfloat[]{\includegraphics[width=0.45 \textwidth]{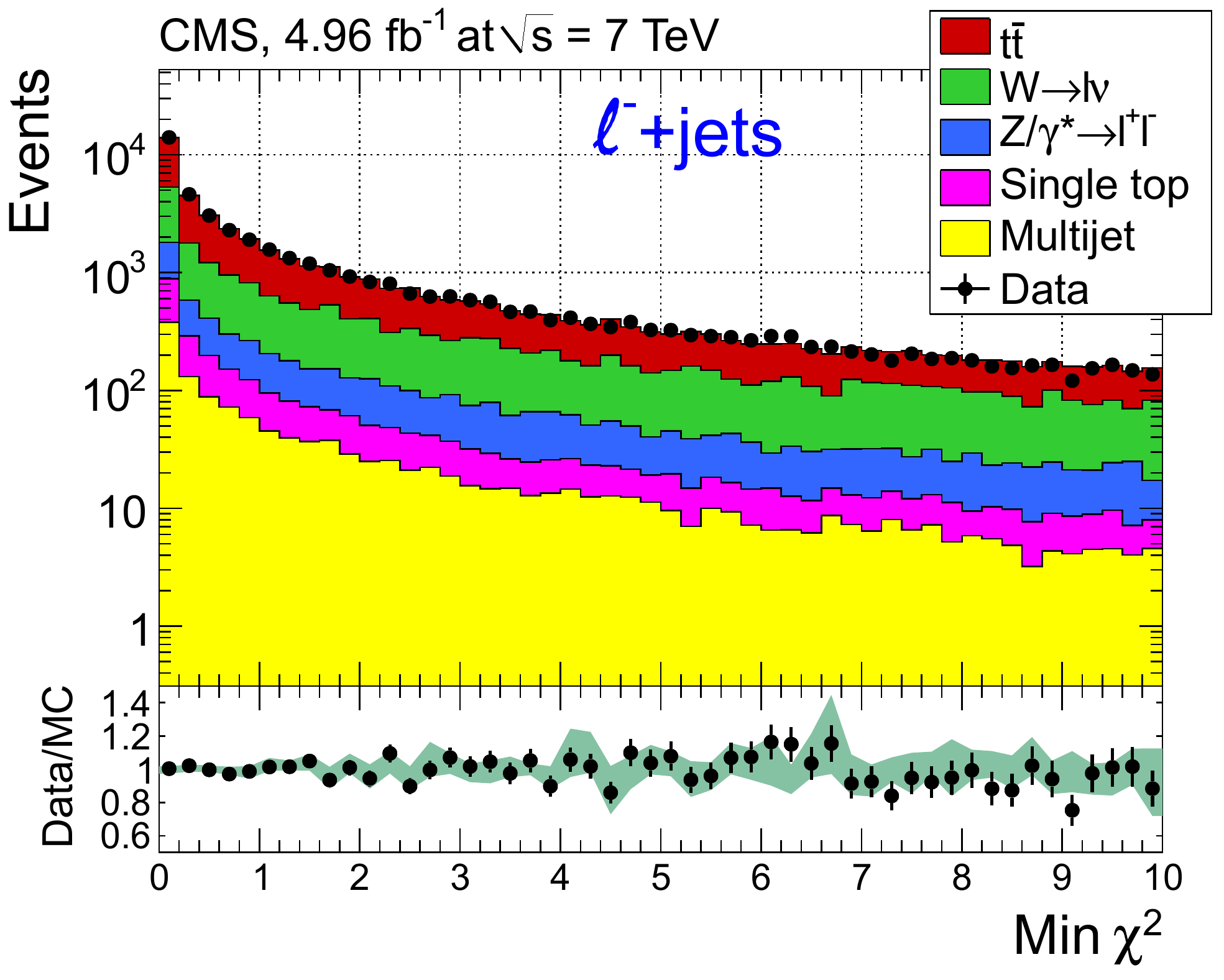} \label{fig:dataMCkinfit4}}
  \caption{
  Panels \protect\subref{fig:dataMCkinfit1} and \protect\subref{fig:dataMCkinfit2} show the distributions in fitted top-quark mass for the smallest fit-$\chi^2$ values, which are given in \protect\subref{fig:dataMCkinfit3} and \protect\subref{fig:dataMCkinfit4}, for the kinematic fits for $\ell^+$+jets and $\ell^-$+jets events, respectively. The simulation is normalized to the number of events observed in data. The last bins include the contributions from overflow. The ratio of the number of observed events in data to the number of events expected from simulation is shown at the bottom. The error band corresponds to the systematic uncertainties related to jet energy scale, jet energy resolution, background estimation and modeling of pileup.
  \label{fig:dataMCkinfit}
  }
\end{figure*}

The fitted values of the top-quark mass $m_i$, the uncertainty on the mass $\sigma_i$ and the $\chi^2_{i}$, obtained for each combination of jets $i$, are used as input to the Ideogram method. A comparison of these variables between data and simulation is given in Fig.~\ref{fig:dataMCkinfit}, for just the jet combination with the smallest $\chi^2$ in each event. Good agreement is observed between data and simulation.

\section{The Ideogram method} \label{sec:ideogram}

In the Ideogram method, the mass of the top quark is measured using a likelihood defined as a function of $m_{\text{t}}$, as follows:
\begin{equation} \label{eq:eventLikel}
  \mathcal{L}_{\rm event} (x;y \mid m_{\text{t}}) = f_{\ttbar} ~ P_{\ttbar}(x;y \mid m_{\text{t}}) + ( 1 - f_{\ttbar} ) P_{\rm bkg} (x).
\end{equation}
This equation expresses the likelihood for any mass $m_{\text{t}}$ in terms of a sum of probability densities from \ttbar and background components. The fraction $f_{\ttbar}$ of the \ttbar component is taken from Table~\ref{table:eventCountsSplitted}. The functions $P_{\ttbar}(x;y \mid m_{\text{t}})$ and $P_{\rm bkg}(x)$ depend on the observables $x$, respectively for the \ttbar and for the background hypotheses, where $x$ includes the number of b-tagged jets $n_{\rm b}$, the lepton charge $q^{\ell}$, and the $m_i$ for each combination of jets $i$ in the event. The quantities $y$ represent the values of the parameters $\sigma_i$ and $\chi^2_{i}$ from the kinematic fits, and are used to parametrize $P_{\ttbar}$, as shown in Eq.~(\ref{eq:signalLikel}) below. The number of b-tagged jets is obtained using the Simple Secondary-Vertex High-Efficiency algorithm (SSVHE)~\cite{CMS-PAS-BTV-11-003} at its "medium" working point.
It is assumed that the background probability $P_{\rm bkg}$ can be described just by the probability density for the main background from W+jets. This is acceptable, as the contributions from other backgrounds are expected to be small, and their probability densities differ greatly from that from \ttbar, but have distributions similar to that for W+jets events.

Furthermore, it is assumed that the number of b-tagged jets and the lepton charge are uncorrelated with the mass information in a given event. This means that the signal and background probabilities can be written as the product of a probability to observe $n_{\rm b}$ b-tagged jets, a probability to observe a certain lepton charge $q^{\ell}$, and a probability to observe $x_{\rm mass}$, which represents the set of mass variables $m_i$ in an event:
\begin{equation}
  P_{\ttbar}(x;y \mid m_{\text{t}}) = P_{\ttbar}(n_{\rm b}) \cdot P_{\ttbar}(q^{\ell}) \cdot P_{\ttbar}(x_{\rm mass};y \mid m_{\text{t}});
\end{equation}
\begin{equation}
  P_{\rm bkg} (x) = P_{\rm bkg} (n_{\rm b}) \cdot P_{\rm bkg}(q^{\ell}) \cdot P_{\rm bkg} (x_{\rm mass}).
\end{equation}
These probability densities for the number of b-tagged jets and lepton charge for signal, $P_{\ttbar}(n_{\rm b})$ and $P_{\ttbar}(q^{\ell})$, and for background, $P_{\rm bkg} (n_{\rm b})$ and $P_{\rm bkg}(q^{\ell})$, are taken from simulation. The reason for including b-tagging at this point is to reduce the impact from non-\ttbar background, while the reason for considering the probability distributions for lepton charge in background is to account for the dependence of W+jets and single-top events on the charge of the lepton.

The \ttbar probability for each event
contains two terms, one representing the probability that a jet combination has the correct jet-to-quark assignment, and the second term expressing the probability that a jet combination has a wrong jet-to-quark assignment, which, summed over all possibilities $i$ in each event, yields:
\begin{equation} \label{eq:signalLikel}
  P_{\ttbar}(x_{\rm mass};y \mid m_{\text{t}}) = \sum_{i=1}^{12} w_i \left( f_{\rm gc} \int
  \mathrm{d}m' G(m_i \mid m' , \sigma_i )
  B ( m' \mid m_{\text{t}} , \Gamma_t ) + ( 1 - f_{\rm gc} ) W ( m_i \mid m_{\text{t}} ) \right).
\end{equation}
The parameter $f_{\rm gc}$ reflects the probability that the jet combination with highest weight $w_i$ (defined below) corresponds to the correct jet-parton matching, as obtained from \ttbar simulation, separately for events with $n_{\rm b} =$ 0, 1, and $> 1$.
The probability for the correct jet combination is defined by the convolution in Eq.~(\ref{eq:signalLikel}) of a Gaussian resolution function $G(m_i \mid m' , \sigma_i )$ and a relativistic Breit-Wigner distribution $B ( m' \mid m_{\text{t}} , \Gamma_t )$. The width of the top quark $\Gamma_t$ is fixed to 2\GeV. The Gaussian function describes the mass resolution for each jet combination. It is centered at the Breit-Wigner-distributed value of the top-quark mass ($m'$) and has a standard deviation equal to the uncertainty on the fitted top-quark mass ($\sigma_i$). If the smallest $\chi^2_{i}$ in an event ($\chi^2_{\rm min}$) is larger than the value of $n_{\rm dof}$, all the $\sigma_i$ for the event are scaled up by a factor $\sqrt{ \chi^2_{\rm min} / n_{\rm dof} }$.
The symbol $W ( m_i \mid m_{\text{t}} )$ in Eq.~(\ref{eq:signalLikel}) represents the probabilities for the wrong jet combinations, which are parametrized using analytic functions fitted to the mass distribution of jet combinations from simulated \ttbar events known to have wrong jet-to-quark assignments.

The probability for \ttbar signal of Eq.~(\ref{eq:signalLikel}) is calculated as a sum over all fitted jet combinations with $\chi^2_{\rm min} < 10$, each weighted by:
\begin{equation} \label{eq:weight}
  w_i = \exp{ \left( - \frac{1}{2} ~ \chi^2_i \right) } w_{\rm b}.
\end{equation}
The first factor above represents the likelihood for the kinematic fit with that combination of jets, while the second factor reflects the degree of compatibility with the observed b-tagging assignments:
\begin{equation}
  w_{\rm b} = \prod_{j} p^j,
\end{equation}
where the index $j$ runs over all jets considered in the fit, and the probabilities $p^j$ equal $\varepsilon_l$, $(1 - \varepsilon_l)$, $\varepsilon_b$, or $(1 - \varepsilon_b)$, depending on the flavor assigned to each jet, and whether the jet is b-tagged. The b-tag efficiency ($\varepsilon_b$) is $60.6 \pm 2.5\%$, and is calculated from \ttbar simulation using the scale factors between data and simulation and the corresponding uncertainties from Ref.~\cite{CMS-PAS-BTV-11-003}. The rate for tagging light-flavor jets ($\varepsilon_l$) is taken from Ref.~\cite{CMS-PAS-BTV-11-001} and equals $1.4 \pm 0.3\%$ for jets with $50 < \PT < 80$\GeV. The individual weights $w_i$ are normalized to sum to unity for each event.

The background probability $P_{\rm bkg}$ in Eq.~(\ref{eq:eventLikel}) does not depend on the mass of the top quark, and has only minimal dependence on the jet-quark assignments. The distribution is therefore defined by the mean of the combined distributions of all solutions for $m_i$ in simulated W+jets events, and fitted to an analytical function.

The combined likelihood for the full event sample is calculated as the product of the individual event likelihoods for all selected events. The fitted top-quark mass and its statistical uncertainty are extracted from this combined likelihood.
While $f_{\ttbar}$ can be treated as a free parameter of the fit~\cite{Abazov:2007rk}, in this analysis it is fixed to the expected value (cf.\ Table~\ref{table:eventCountsSplitted}) and the uncertainty on the signal fraction is taken into account as a systematic uncertainty of the method.

\section{Calibration of individual mass measurements} \label{sec:calibration}

The likelihood defined in Eq.~(\ref{eq:eventLikel}) for each event corresponds to a simplified model for the ensuing analysis, which means that the resulting combined likelihood reflects an approximate quantity. To correct for possible bias in the estimated mass or in the estimate of statistical uncertainty, a calibration of the procedure is performed using pseudo-experiments. In these pseudo-experiments, events are picked randomly from samples of simulated events representing the major contributing processes in Table~\ref{table:eventCountsSplitted}, implementing Poisson fluctuations around the respective means as expected in true data. The distributions for multijet events are modeled using control samples of data, as described in Sec.~\ref{sec:selection}. For \ttbar signal, nine samples of simulated events are generated for top-quark masses between 161.5 and 184.5\GeV. The calibration is performed for the accepted inclusive ($> 3$ jets) samples of $\ell$+jets events.

\begin{figure*}[htb]
  \centering
  \subfloat[]{\includegraphics[angle=90,width=0.49 \textwidth]{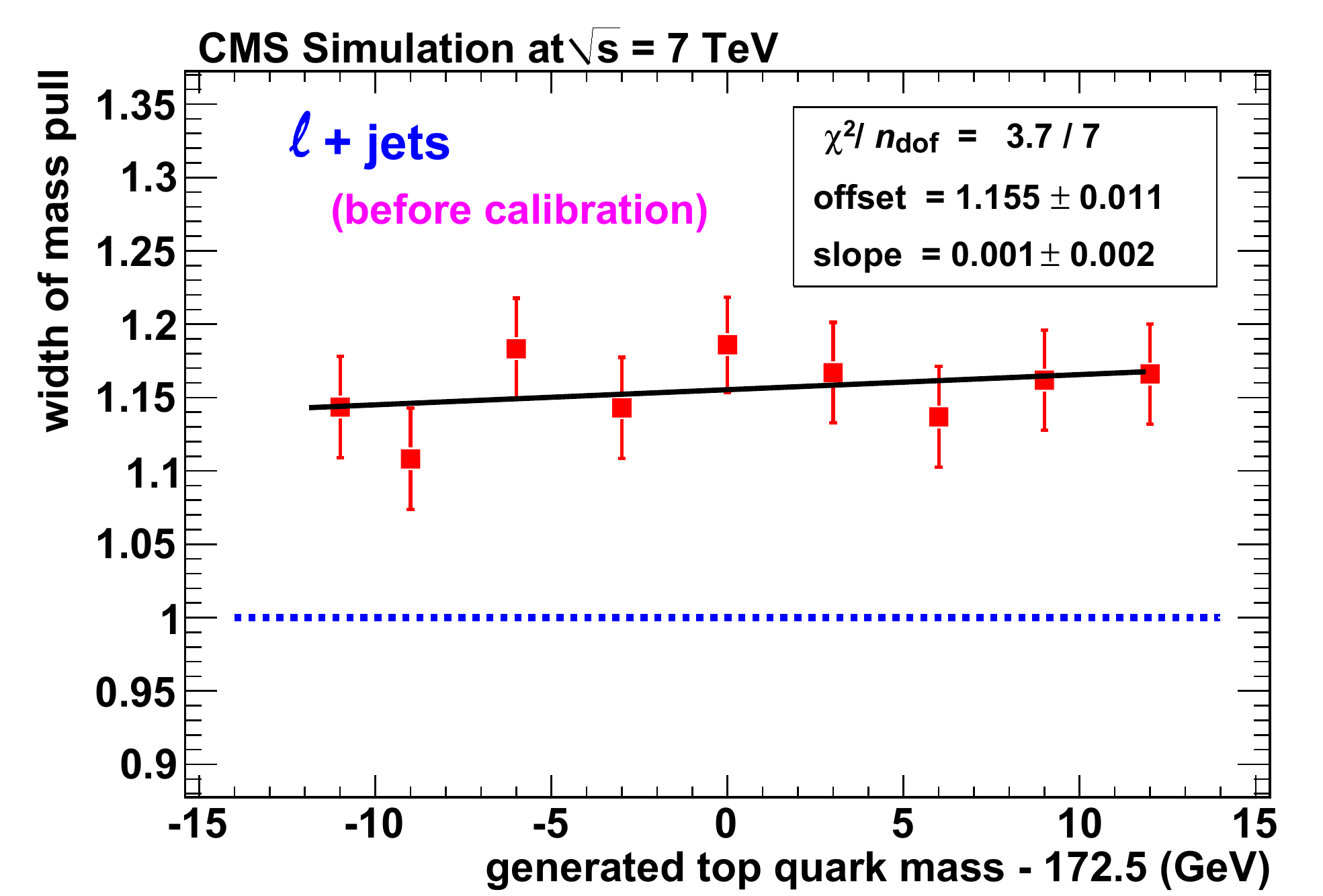} \label{fig:calibPullBias1}}
  \subfloat[]{\includegraphics[angle=90,width=0.49 \textwidth]{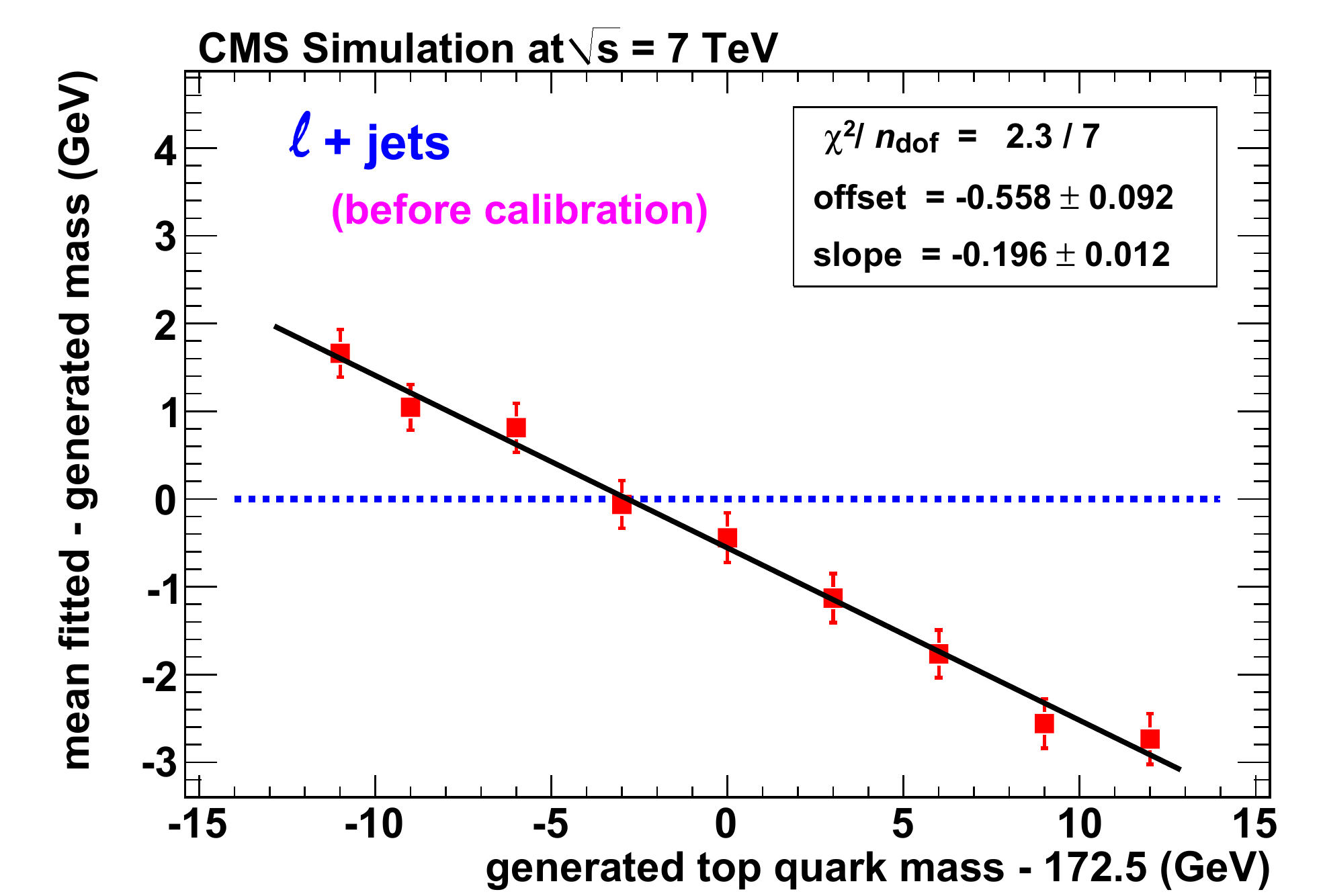} \label{fig:calibPullBias2}}
  \caption{
  \protect\subref{fig:calibPullBias1} Width of the pull distribution and \protect\subref{fig:calibPullBias2} bias on the extracted top-quark mass, as a function of generated top-quark mass for $\ell$+jets events simulated in ensembles of pseudo-experiments.
\label{fig:calibPullBias}
}
\end{figure*}

\begin{figure*}[htb]
  \centering
  \subfloat[]{\includegraphics[angle=90,width=0.49 \textwidth]{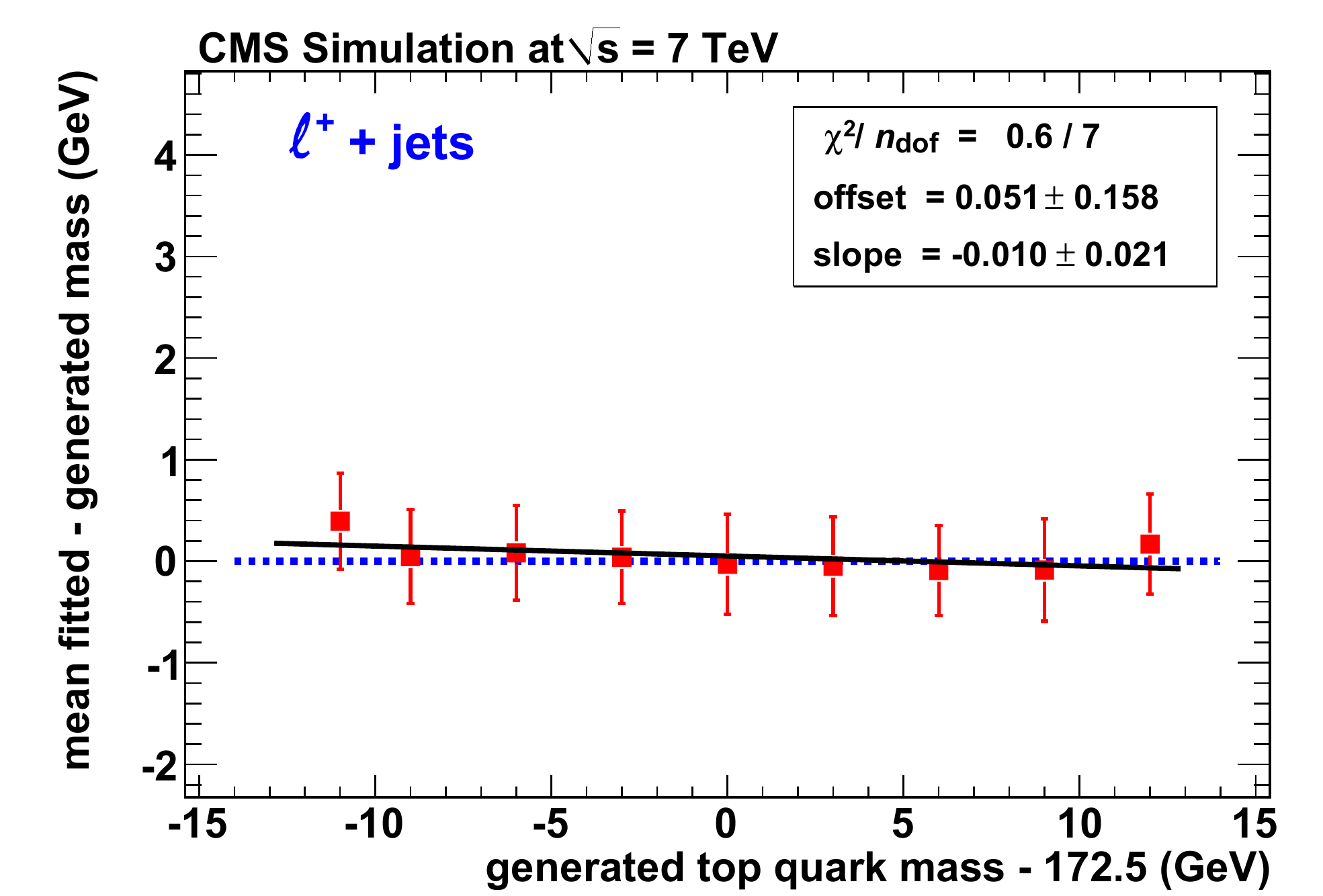} \label{fig:calibratedBias1}}
  \subfloat[]{\includegraphics[angle=90,width=0.49 \textwidth]{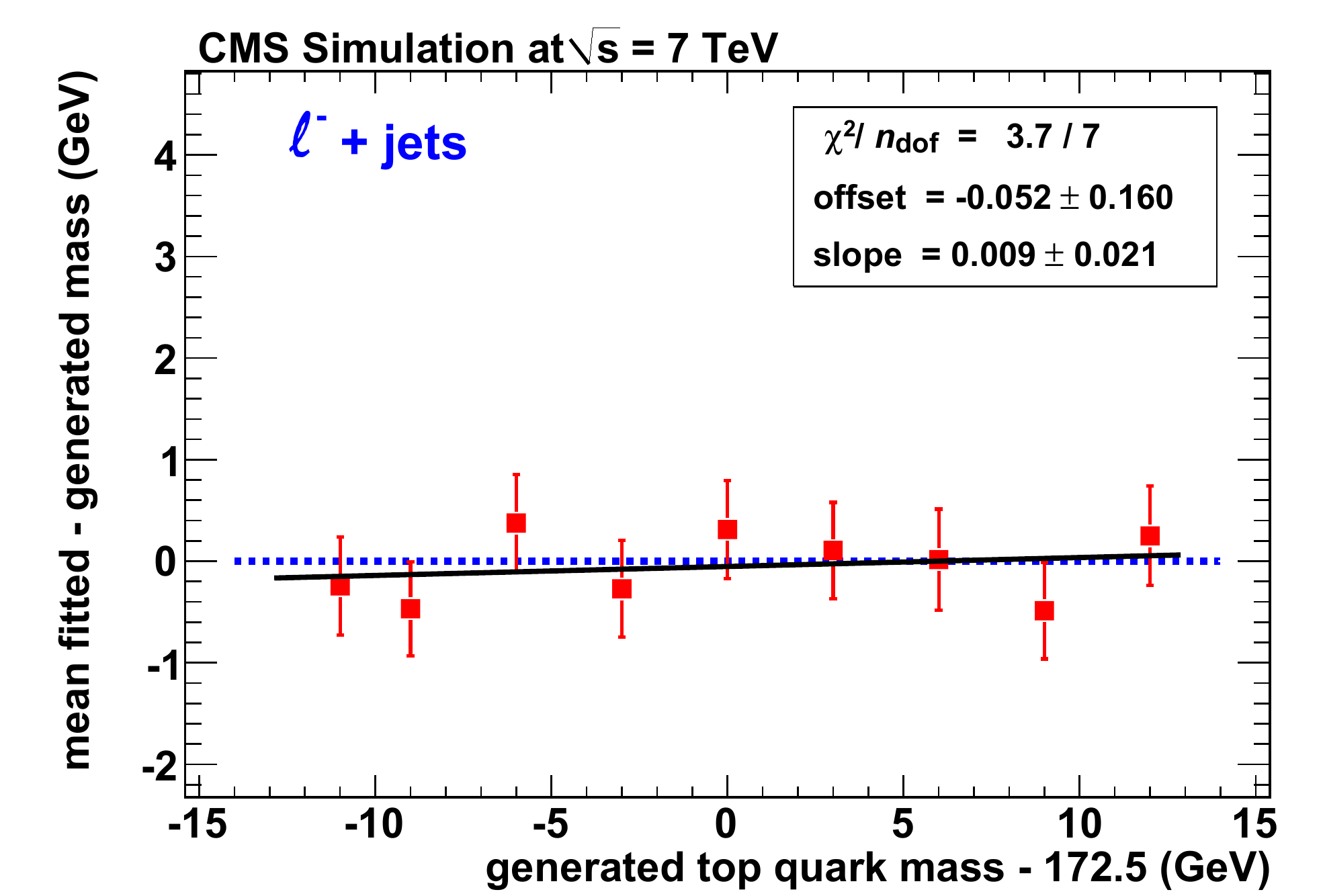} \label{fig:calibratedBias2}} \\ \vspace{0.3cm}
  \subfloat[]{\includegraphics[angle=90,width=0.49 \textwidth]{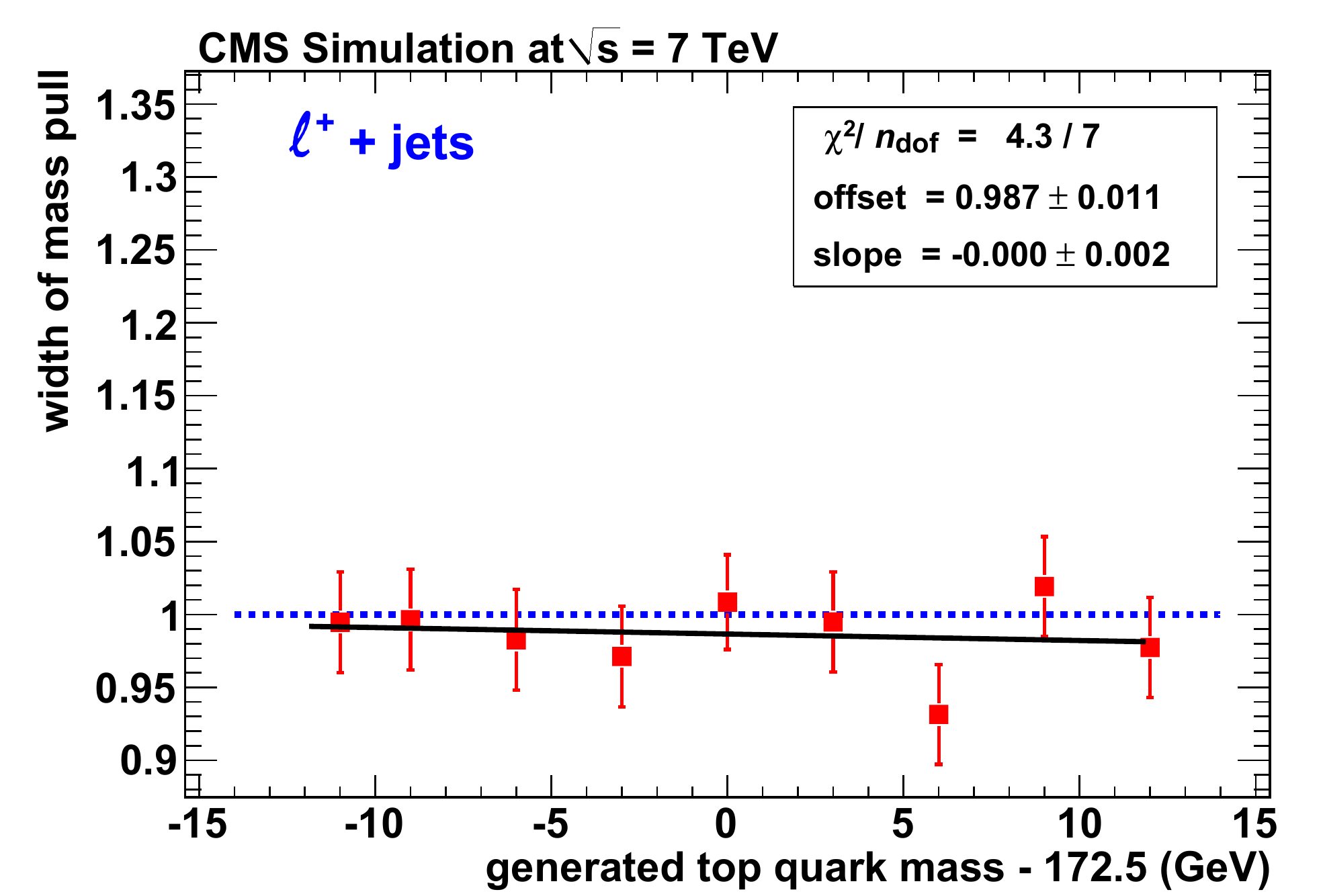} \label{fig:calibratedBias3}}
  \subfloat[]{\includegraphics[angle=90,width=0.49 \textwidth]{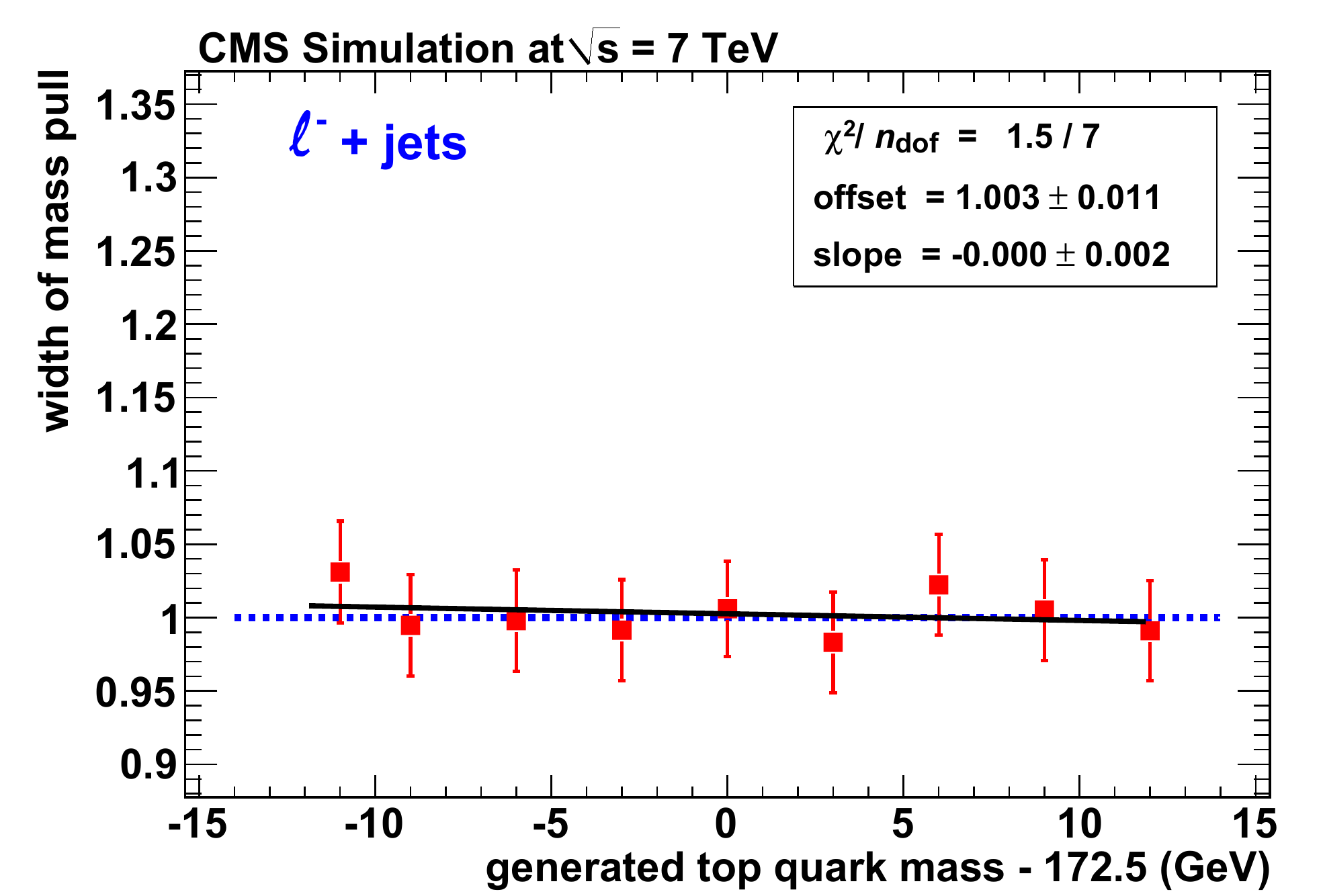} \label{fig:calibratedBias4}}
  \caption{
  \label{fig:calibratedBias}
  }
\end{figure*}

The widths of pull distributions and the bias on the estimated top-quark mass as a function of generated mass are shown for the combined $\ell^+$ and $\ell^-$ events in Fig.~\ref{fig:calibPullBias}. The pull is defined as the standard deviation of a Gaussian function fitted to the distribution of $(m_j - \langle m \rangle ) / \sigma_j$, where $m_j$ is the estimated top-quark mass in each pseudo-experiment, $\sigma_j$ its estimated statistical uncertainty and $\langle m \rangle$ the mean of the estimated top-quark masses over all pseudo-experiments at a fixed input mass. 
Since the standard deviation of the pull distribution appears to be slightly larger than unity, the statistical uncertainties on the final mass measurement are scaled up by that discrepancy ($\approx 16\%$). Also, as seen from Fig.~\ref{fig:calibPullBias}, the bias on the estimated top-quark mass depends linearly on the generated top-quark mass. Although these biases are within 2\GeV for most of the range of interest, they are corrected using the fitted linear calibration given in the figure. 
The bias on the estimated top-quark mass and width of the pull as a function of generated mass are shown
separately for $\ell^+$+jets and $\ell^-$+jets events in Fig.~\ref{fig:calibratedBias} (after implementing the inclusive $\ell$+jets calibration from Fig.~\ref{fig:calibPullBias}). The results show that, within statistical precision, the separate $\ell^+$+jets and $\ell^-$+jets events do not require additional independent corrections.

\section{Measurement of the \texorpdfstring{$\text{t} - \overline{\text{t}}$}{t t-bar} mass difference} \label{sec:diffMeasurement}

For the final measurement of the mass difference, we apply the analysis separately to $\ell^-$+jets events and to $\ell^+$+jets events, and take the difference of the two extracted values.
In the \Pgm+jets channel, the individual measurements yield a mass difference of:
\begin{equation*}
  \Delta m_{\text{t}} = 0.13 \pm 0.61 ~\stat\GeV,
\end{equation*}
and in the \Pe+jets channel:
\begin{equation*}
  \Delta m_{\text{t}} = -1.28 \pm 0.70 ~\stat\GeV,
\end{equation*}
and when the method is applied to the combined \Pe+jets and \Pgm+jets samples:
\begin{equation*}
  \Delta m_{\text{t}} = -0.44 \pm 0.46 ~\stat\GeV.
\end{equation*}

The results for $\Delta m_{\text{t}}$ are compatible with the expectation from the hypothesis of CPT symmetry, which forbids a mass difference between the top quark and the antitop quark. Also, the average fitted top-quark mass is found to be $m_{\text{t}} = 173.36 \pm 0.23 ~\stat\GeV$, which is in agreement with previous measurements of $m_{\text{t}}$~\cite{CMS-PAPERS-TOP-11-002,ATLAS-TOPMASS,TopMassCDF,TopMassD0}, even ignoring systematic uncertainties.

\section{Systematic uncertainties} \label{sec:systematics}

Many of the systematic uncertainties relevant for the absolute measurement of $m_{\text{t}}$, such as the calibration of the overall jet energy scale, are reduced in the context of this measurement, as such systematic effects tend to alter the measured properties of top and antitop quarks in a similar and correlated manner. Several other sources of systematic uncertainty on the modeling of physical processes evaluated in absolute $m_{\text{t}}$ measurements are not expected to affect the measurement of $\Delta m_{\text{t}}$. These include modeling of hadronization, the underlying event, initial and final-state radiation, changes in factorization and renormalization scales, and the matching of partons to parton showers, and are not considered further in this analysis.

Systematic uncertainties for other effects considered in the measurement of $m_{\text{t}}$ are included together with additional sources potentially relevant for $\Delta m_{\text{t}}$, such as lepton-charge identification and a possible difference in jet response to b and $\overline{\rm b}$ quarks. These are listed in Table~\ref{table:Systematics}, and described in greater detail below. In all cases, the effects are evaluated using simulated event samples, by comparing the nominal sample to one where the systematic effect is varied by $\pm 1$ standard deviation. Statistical uncertainties on the observed mass shifts are evaluated using the resampling technique of Ref.~\cite{Jackknife}, and are listed in Table~\ref{table:Systematics}. For most systematic uncertainties, the statistical significance of the observed shift in $\Delta m_{\text{t}}$ is small. We therefore quote the observed shift as a systematic uncertainty when it is larger than the statistical uncertainty, and otherwise we quote just the statistical uncertainty. The total systematic uncertainty is taken to be the quadratic sum of the values quoted for each source.

\begin{table}[htb]
  \begin{center}
    \caption{Overview of systematic uncertainties on $\Delta m_{\text{t}}$. The total is defined by adding in quadrature the contributions from all sources, by choosing for each the larger of the estimated shift or its statistical uncertainty, as indicated by the bold script. \label{table:Systematics}}
    \begin{tabular}{lc}
      \hline \hline
      Source	                                    & Estimated effect (GeV) \\ \hline
      Jet energy scale		                        & $ 0.04 \pm {\bf 0.08} $ \\
      Jet energy resolution	                      & $ 0.04 \pm {\bf 0.06} $ \\
      b vs. $\rm \overline{\rm b}$ jet response        & $ {\bf 0.10} \pm  0.10 $  \\
      Signal fraction                             & $ {\bf 0.02} \pm 0.01 $ \\
      Difference in \PWp/\PWm production          & $ {\bf 0.014}\pm 0.002 $\\
      Background composition	                    & $ {\bf 0.09} \pm 0.07 $  \\
      Pileup                                      & $ {\bf 0.10} \pm 0.05 $  \\
      b-tagging efficiency                        & $ {\bf 0.03} \pm 0.02 $ \\
      b vs. $\rm \overline{\rm b}$ tagging efficiency  & $ {\bf 0.08} \pm 0.03 $ \\
      Method calibration                          & $ 0.11 \pm {\bf 0.14}  $\\
      Parton distribution functions               & $ {\bf 0.088} $ \\ \hline
      Total & { 0.27}\\ \hline \hline
    \end{tabular}
  \end{center}
\end{table}

\begin{description}
  \item[Overall jet energy scale.] The uncertainty related to the overall jet energy scale is estimated by changing the energy of all jets within their $\PT$ and $\eta$-dependent uncertainties. This uncertainty contains contributions from the uncertainty on pileup and flavor dependence of jets. These are measured using 2010 data~\cite{JESpaper}, and have several additional uncertainties added in quadrature: 1.5\% from a mismatch in absolute energy between Z+jets and $\gamma$+jets events, and 1.5\% for jets with $|\eta|>1.3$ from an $\eta$-dependence on the relative scale. Because top and antitop quarks at the LHC are produced with slightly different distributions in rapidity, an $\eta$-dependence for jet response can lead to a small residual effect on $\Delta m_{\text{t}}$. While the average extracted top-quark mass shifts by as much as $\pm 2.3$\GeV, the observed effect on $\Delta m_{\rm t}$ is only $0.04 \pm 0.08 \stat$\GeV. We quote the statistical precision on the shift as a systematic uncertainty of the measurement.
  \item[Jet energy resolution.] Previous measurements of jet energy resolution in data have indicated that it is 10\% worse than in the simulation~\cite{JESpaper}. The resolution in simulated events used for calibration is therefore degraded accordingly. The uncertainty on this 10\% depends on $\eta$, and equals $\pm 10\%$ for jets within $|\eta| < 1.5$, $\pm 15\%$ for jets within $1.5 < |\eta| < 2.0$, and $\pm 20\%$ for jets with $|\eta| > 2.0$. Based on generated parton energies, the resolution for each jet is scaled up and down within these uncertainties. Half of the difference between such up and down changes yields a $-0.04 \pm 0.06 \stat$\GeV difference in $\Delta m_{\rm t}$. While this is expected to be the same for the $m_{\text{t}}$ and $m_{\overline{\text{t}}}$ measurements, a residual effect is possible through the asymmetry in the composition of the background. This possibility is included as a systematic uncertainty on $\Delta m_{\rm t}$.
  \item[Jet energy scale for b and $\bf \overline{b}$.] A dedicated study is performed to assess the jet response for b and $\overline{\rm b}$ jets, by comparing the reconstructed jet $\PT$ to the original parton $\PT$ in \ttbar simulation as a function of jet $\eta$ and $\PT$.
The \PYTHIA simulation describes differences in the fragmentation of b and $\overline{\rm b}$ jets, including B-$\overline{\rm B}$ oscillations, and the CMS detector simulation includes differences in calorimeter response for K$^+$ and K$^-$ particles. As the PF algorithm reconstructs charged hadrons using tracks when available, the impact of such differences in calorimeter response on the reconstructed jet energy is expected to be small. On average, the ratio of b to $\overline{\rm b}$ response is found to be $0.999 \pm 0.001$, compatible with unity. In principle, the calibration would correct for such differences, albeit with limited statistical precision (see below). Nevertheless, we quote 100\% of the corresponding shift of $0.10$\GeV as a systematic uncertainty on $\Delta m_{\rm t}$.
  \item[Signal fraction.] An incorrect fraction of \ttbar signal events in the simulation would, in principle, bias the calibration procedure. Changing the relative signal fraction by $\pm20$\%, while keeping the background composition fixed, yields an effect of $\mp~ 0.02 \pm 0.01$\GeV on $\Delta m_{\rm t}$.
  \item[Difference in \PWp/\PWm~production.] The difference in production cross sections of \PWp ~and~ ${\rm W}^{-}$ bosons in pp collisions leads to different levels of \PW+jets background and different background composition in $\ell^+$+jets and $\ell^-$+jets channels. This can affect the calibration procedure and lead to a small bias in $\Delta m_{\rm t}$. The measured inclusive \PWp/\PWm~ratio is in agreement with theoretical prediction within a precision of 3.5\%~\cite{WZ}, and has been studied by CMS as a function of pseudorapidity~\cite{ChargeAsymmetry}. Varying the $\ell^+$ and $\ell^-$ backgrounds by 2\% in opposite directions, thereby affecting the relative ratio of \PWp~and \PWm~by 4\%, changes $\Delta m_{\rm t}$ by $-14 \pm 2$~MeV, which is quoted as the systematic uncertainty resulting from the difference due to unequal yields of \PWp~and \PWm.
  \item[Background composition.] To evaluate any residual effects related to distributions and composition of the background, we investigate the effect of removing completely each source of background from the calibration procedure, while keeping the signal fraction constant. We quote 30\% of the total shift observed in $\Delta m_{\rm t}$ when we remove W+jets ($-0.26 \pm 0.20$\GeV), Z+jets ($0.05 \pm 0.04$\GeV) and single top-quark production ($0.05 \pm 0.02$\GeV), and 100\% for the background from multijet events ($-2 \pm 5$~MeV). For each contribution, we take the larger of the observed shift in $\Delta m_{\rm t}$ or its statistical uncertainty and add the four sources in quadrature.
  \item[Pileup.] The simulated events used in this analysis contain contributions from pileup, and are reweighted to match the estimated dependence of pileup on instantaneous luminosity. The systematic uncertainty is estimated by changing the mean value of the number of interactions by $\pm 0.6$, and taking the average of the two shifts in $\Delta m_{\rm t}$ as the systematic uncertainty. This covers the uncertainty in the modeling of pileup as well as the uncertainty on the calculation of event weights. The uncertainty on the weights is dominated by uncertainties on the total inelastic cross section and on the measured luminosity, both of which are used in the reweighting. To further investigate any additional effects related to high pileup conditions during high-luminosity running, the measured values are examined as a function of the number of reconstructed primary vertices in Fig.~\ref{fig:PU}. No adverse effects are observed, and the results for $\Delta m_{\rm t}$ are stable and statistically compatible with no dependence on the number of pileup events in the data.
\begin{figure}[htb]
  \centering
  \includegraphics[width=0.49 \textwidth]{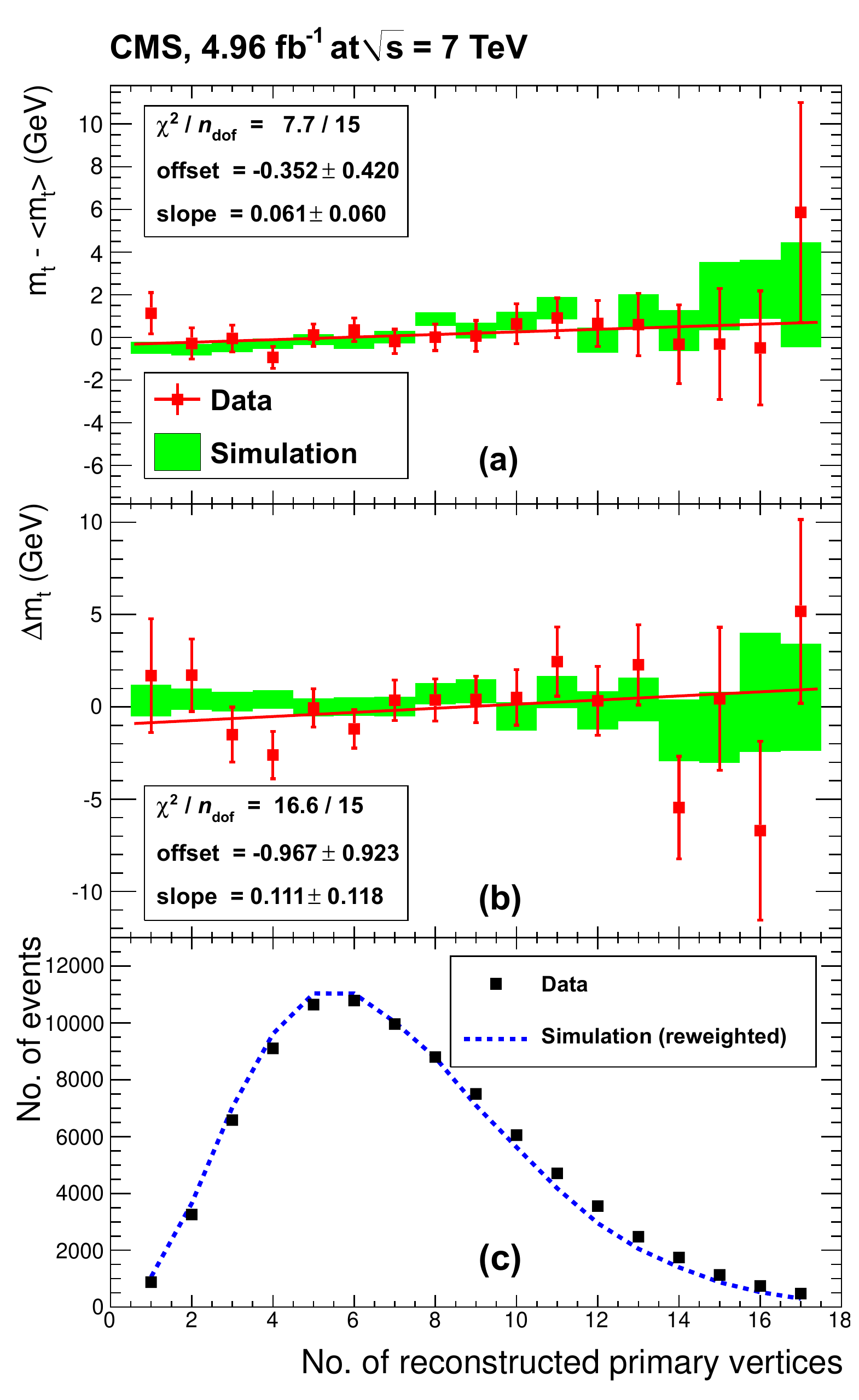}
  \caption{Difference between (a) the measured $m_{\text{t}}$ in each bin and the average $m_{\text{t}}$, and (b) the value of $\Delta m_{\text{t}}$ both for data and for simulation, as a function of the number of reconstructed collision vertices. The results of fitting the data with a linear function are also shown. (c) Distribution for the number of reconstructed collision vertices in data and in simulation, after reweighting for pileup.\label{fig:PU}}
\end{figure}
  \item[B-tagging efficiency.] A shift in b-tagging efficiency can affect the impact of background processes on the Ideogram method. In particular, a difference in distributions of positively and negatively charged particles in the background can affect $\Delta m_{\rm t}$. As indicated previously, the SSVHE tagger is used primarily at its "medium" working point. To quantify the impact of b-tagging efficiency, we vary the threshold defining the working point, thereby producing a relative change in efficiency of $\pm 4$\%~\cite{CMS-PAS-BTV-11-003}. The changes are applied in the same direction or opposite direction for the $\ell^+$+jets and $\ell^-$+jets samples, and the corresponding shifts on $\Delta m_{\text{t}}$ of 0.03\GeV (b-tagging efficiency) and 0.08\GeV (b versus $\rm \overline{\rm b}$ tagging efficiency) are quoted as systematic uncertainties.
  \item[Misassignment of lepton-charge.] The leptons are used only in triggering and splitting the data into $\ell^+$+jets and $\ell^-$+jets events, but not in mass reconstruction. A wrong assignment of charge can affect the calibration of $\Delta m_{\text{t}}$ in a way that is not recovered in the overall procedure. It can also lead to a cross-contamination of the two event samples, which can bias or dilute the measurement. The rate of charge misassignment in muon reconstruction is measured with cosmic muons~\cite{MuonPerformance} and collision data~\cite{ChargeAsymmetry} to be of the order of 10$^{-3}$ to 10$^{-2}$\% in the transverse momentum range of this measurement. For electrons, the rate of charge misidentification ranges from 0.1\% to 0.4\%, depending on pseudorapidity~\cite{ChargeAsymmetry}. This means that the systematic uncertainty from charge misassignment is below 1\% of the measured $\Delta m_{\text{t}}$ value, which is negligible and is therefore ignored.
  \item[Trigger.] The trigger requires the presence of a lepton and at least three jets. As the lepton is not used in mass reconstruction, no systematic effect is expected from any mismodeling of the lepton trigger efficiency or $\PT$ threshold. The requirement of three jets in the trigger is highly efficient for events with 4 jets with $\PT > 30$\GeV. Any effect on kinematic distributions of the jets in selected events is therefore estimated to be small, and expected to affect the $m_{\text{t}}$ and $m_{\overline{\text{t}}}$ measurements equally. No uncertainty is quoted therefore for this source.
  \item[Method calibration.] The effect is evaluated for simulated \ttbar events at a mass of 172.5\GeV, showing a difference in mass bias between $\ell^+$+jets and $\ell^-$+jets of $-0.11 \pm 0.14$\GeV, which is statistically compatible with no effect.
This confirms our expectation that there is no known effect in simulation that would lead to a difference in mass calibration between the two channels. Based on this observation, the combined $\ell$+jets calibration is applied both in the $\ell^{+}$+jets and the $\ell^{-}$+jets channel. The statistical uncertainty on the calibration of the mass difference is quoted as a systematic uncertainty of 0.14\GeV.
  \item[Parton distribution functions.] The choice of parton distribution functions (PDF) can affect $\Delta m_{\text{t}}$, as they determine, for example, the difference in production of \PWp and \PWm, which is the dominant source of background. The simulated samples are generated using the CTEQ 6.6 PDF~\cite{CTEQ}, for which the uncertainties can be described by 22 independent parameters. Up and down changes in these parameters result in 22 accompanying PDF possibilities. Using a simulated sample of \ttbar and background events, reweighted according to the deviation of each PDF from its original form, the sum of the larger shift (``up'' or ``down'') for each change in PDF is taken in quadrature, to define an estimated combined uncertainty on $\Delta m_{\text{t}}$ of 88\MeV.
\end{description}

\section{Summary} \label{sec:conclusion}

The mass difference between the top quark and the antitop quark, $\Delta m_{\text{t}} = m_{\text{t}} - m_{\overline{\text{t}}}$, is measured with the Ideogram method using the $\ell$+jets \ttbar event sample collected by the CMS experiment, corresponding to an integrated luminosity of $4.96 \pm 0.11$~\fbinv. This yields the result:
\begin{equation*}
  \Delta m_{\text{t}} = -0.44 \pm 0.46 ~\stat \pm 0.27 ~\syst\GeV
\end{equation*}
The measured value is in agreement with the consequence of CPT invariance, which requires no mass difference between the top and antitop quarks. This is more precise by at least a factor three than any of the previous measurements.

\section*{Acknowledgments}
We congratulate our colleagues in the CERN accelerator departments for the excellent performance of the LHC machine, thank the technical and administrative staffs at CERN and other CMS institutes for their assistance, and acknowledge support from: FMSR (Austria); FNRS and FWO (Belgium); CNPq, CAPES, FAPERJ, and FAPESP (Brazil); MES (Bulgaria); CERN; CAS, MoST, and NSFC (China); COLCIENCIAS (Colombia); MSES (Croatia); RPF (Cyprus); Academy of Sciences and NICPB (Estonia); Academy of Finland, MEC, and HIP (Finland); CEA and CNRS/IN2P3 (France); BMBF, DFG, and HGF (Germany); GSRT (Greece); OTKA and NKTH (Hungary); DAE and DST (India); IPM (Iran); SFI (Ireland); INFN (Italy); NRF and WCU (Korea); LAS (Lithuania); CINVESTAV, CONACYT, SEP, and UASLP-FAI (Mexico); MSI (New Zealand); PAEC (Pakistan); SCSR (Poland); FCT (Portugal); JINR (Armenia, Belarus, Georgia, Ukraine, Uzbekistan); MON, RosAtom, RAS and RFBR (Russia); MSTD (Serbia); MICINN and CPAN (Spain); Swiss Funding Agencies (Switzerland); NSC (Taipei); TUBITAK and TAEK (Turkey); STFC (United Kingdom); DOE and NSF (USA).

\bibliography{auto_generated}   

\providecommand{\href}[2]{#2}\begingroup\raggedright\begin{thebibliography}{10}%
\makeatletter
\providecommand{\hrefCMSnoop }[0]{\@secondoftwo}%
\makeatother
\providecommand{\doi}{\texttt{doi:}\begingroup \urlstyle{tt}\Url}

\bibitem{PDG}
\hrefCMSnoop {} {K.~Nakamura {et~al.}, ``The Review of Particle Physics'',}
  \textit{ J. Phys. G} \textbf{ 37} (2010) 075021,
  \href{http://dx.doi.org/10.1088/0954-3899/37/7A/075021}{\doi{10.1088/0954-3899/37/7A/075021}}.

\bibitem{D0massDiffOrg}
\hrefCMSnoop {} {{ D0} Collaboration, ``Direct measurement of the mass
  difference between top and antitop quarks'',} \textit{ Phys. Rev. Lett.}
  \textbf{ 103} (2009) 132001,
  \href{http://dx.doi.org/10.1103/PhysRevLett.103.132001}{\doi{10.1103/PhysRevLett.103.132001}},
  \href{http://www.arXiv.org/abs/0906.1172}{\texttt{ arXiv:0906.1172}}.

\bibitem{CDFmassDiff}
\hrefCMSnoop {} {{ CDF} Collaboration, ``Measurement of the mass difference
  between $t$ and $\bar{t}$ quarks'',} \textit{ Phys. Rev. Lett.} \textbf{ 106}
  (2011) 152001,
  \href{http://dx.doi.org/10.1103/PhysRevLett.106.152001}{\doi{10.1103/PhysRevLett.106.152001}},
  \href{http://www.arXiv.org/abs/1103.2782}{\texttt{ arXiv:1103.2782}}.

\bibitem{D0massDiff}
\hrefCMSnoop {} {{ D0} Collaboration, ``Direct measurement of the mass
  difference between top and antitop quarks'',} \textit{ Phys. Rev. D} \textbf{
  84} (2011) 052005,
  \href{http://dx.doi.org/10.1103/PhysRevD.84.052005}{\doi{10.1103/PhysRevD.84.052005}},
  \href{http://www.arXiv.org/abs/hep-ex/1106.2063}{\texttt{
  arXiv:hep-ex/1106.2063}}.

\bibitem{lhc}
\hrefCMSnoop {} {L.~Evans and P.~{Bryant (editors)}, ``{LHC Machine}'',}
  \textit{ JINST} \textbf{ 03} (2008) S08001,
\href{http://dx.doi.org/10.1088/1748-0221/3/08/S08001}{\doi{10.1088/1748-0221/3/08/S08001}}.

\bibitem{Abdallah:2008xh}
\hrefCMSnoop {} {{ DELPHI} Collaboration, ``{Measurement of the Mass and Width
  of the $W$ Boson in $e^{+}e^{-}$ Collisions at $\sqrt{s}$ = 161 - 209
  GeV}'',} \textit{ Eur. Phys. J. D} \textbf{ 55} (2008) 1,
  \href{http://dx.doi.org/10.1140/epjc/s10052-008-0585-7}{\doi{10.1140/epjc/s10052-008-0585-7}},
\href{http://www.arXiv.org/abs/0803.2534}{\texttt{ arXiv:0803.2534}}.

\bibitem{Abazov:2007rk}
\hrefCMSnoop {} {{ D0} Collaboration, ``{Measurement of the top quark mass in
  the lepton + jets channel using the Ideogram method}'',} \textit{ Phys. Rev.
  D} \textbf{ 75} (2007) 092001,
  \href{http://dx.doi.org/10.1103/PhysRevD.75.092001}{\doi{10.1103/PhysRevD.75.092001}},
\href{http://www.arXiv.org/abs/hep-ex/0702018}{\texttt{ arXiv:hep-ex/0702018}}.

\bibitem{Aaltonen:2006xc}
\hrefCMSnoop {} {{ CDF} Collaboration, ``{Measurement of the top-quark mass in
  all-hadronic decays in $p\bar{p}$ collisions at CDF II}'',} \textit{ Phys.
  Rev. Lett.} \textbf{ 98} (2007) 142001,
  \href{http://dx.doi.org/10.1103/PhysRevLett.98.142001}{\doi{10.1103/PhysRevLett.98.142001}},
\href{http://www.arXiv.org/abs/hep-ex/0612026}{\texttt{ arXiv:hep-ex/0612026}}.

\bibitem{CMS:2008zzk}
\hrefCMSnoop {} {{ {CMS}} Collaboration, ``The {CMS} experiment at the {CERN}
  {LHC}'',} \textit{ JINST} \textbf{ 03} (2008) S08004,
\href{http://dx.doi.org/10.1088/1748-0221/3/08/S08004}{\doi{10.1088/1748-0221/3/08/S08004}}.

\bibitem{MadGraph}
\hrefCMSnoop {} {J.~Alwall, M.~Herquet, F.~Maltoni{ et~al.}, ``MadGraph 5 :
  Going Beyond'',} \textit{ JHEP} \textbf{ 06} (2011) 128,
  \href{http://dx.doi.org/10.1007/JHEP06(2011)128}{\doi{10.1007/JHEP06(2011)128}},
  \href{http://www.arXiv.org/abs/1106.0522}{\texttt{ arXiv:1106.0522}}.

\bibitem{pythia}
\hrefCMSnoop {} {T.~Sj{\"o}strand, S.~Mrenna, and P.~Skands, ``PYTHIA 6.4
  Physics and Manual'',} \textit{ JHEP} \textbf{ 05} (2006) 026,
  \href{http://dx.doi.org/10.1088/1126-6708/2006/05/026}{\doi{10.1088/1126-6708/2006/05/026}},
  \href{http://www.arXiv.org/abs/hep-ph/0603175}{\texttt{
  arXiv:hep-ph/0603175}}.

\bibitem{MLMmatching}
\hrefCMSnoop {} {M.~L. Mangano, M.~Moretti, F.~Piccinini{ et~al.}, ``Matching
  matrix elements and shower evolution for top-quark production in hadronic
  collisions'',} \textit{ JHEP} \textbf{ 01} (2007) 013,
  \href{http://dx.doi.org/10.1088/1126-6708/2007/01/013}{\doi{10.1088/1126-6708/2007/01/013}},
  \href{http://www.arXiv.org/abs/hep-ph/0611129}{\texttt{
  arXiv:hep-ph/0611129}}.

\bibitem{Powheg}
\hrefCMSnoop {} {S.~Frixione, P.~Nason, and C.~Oleari, ``Matching NLO QCD
  computations with parton shower simulations: the \textsc{POWHEG} method'',}
  \textit{ JHEP} \textbf{ 11} (2007) 070,
  \href{http://dx.doi.org/10.1088/1126-6708/2007/11/070}{\doi{10.1088/1126-6708/2007/11/070}},
  \href{http://www.arXiv.org/abs/0709.2092}{\texttt{ arXiv:0709.2092}}.

\bibitem{FEWZ}
\hrefCMSnoop {} {K.~Melnikov and F.~Petriello, ``Electroweak gauge boson
  production at hadron colliders through O($\alpha_s^2$)'',} \textit{ Phys.
  Rev. D} \textbf{ 74} (2006) 114017,
  \href{http://dx.doi.org/10.1103/PhysRevD.74.114017}{\doi{10.1103/PhysRevD.74.114017}},
  \href{http://www.arXiv.org/abs/hep-ph/0609070}{\texttt{
  arXiv:hep-ph/0609070}}.

\bibitem{MCFM}
\hrefCMSnoop {} {J.~M. Campbell and R.~K. Ellis, ``MCFM for the Tevatron and
  the LHC'',} \textit{ Nucl. Phys. B Proc. Suppl.} \textbf{ 205} (2010) 10,
  \href{http://dx.doi.org/10.1016/j.nuclphysbps.2010.08.011}{\doi{10.1016/j.nuclphysbps.2010.08.011}},
  \href{http://www.arXiv.org/abs/1007.3492}{\texttt{ arXiv:1007.3492}}.

\bibitem{Geant4}
\hrefCMSnoop {} {J.~Allison {et~al.}, ``{Geant4} developments and
  applications'',} \textit{ IEEE Trans. Nucl. Sci.} \textbf{ 53} (2006) 270,
\href{http://dx.doi.org/10.1109/TNS.2006.869826}{\doi{10.1109/TNS.2006.869826}}.

\bibitem{CMS-PAS-PFT-10-002}
\href {http://cdsweb.cern.ch/record/1279341} {{ CMS} Collaboration,
  ``Commissioning of the Particle-Flow Reconstruction in Minimum-Bias and Jet
  Events from {\Pp\Pp} Collisions at 7 {TeV}'',} CMS Physics Analysis Summary
  CMS-PAS-PFT-10-002, (2010).

\bibitem{CMS-PAS-MUO-10-002}
\href {http://cdsweb.cern.ch/record/1279140} {{ CMS} Collaboration,
  ``Performance of muon identification in pp collisions at $\sqrt{s}$ = 7
  {TeV}'',} CMS Physics Analysis Summary CMS-PAS-MUO-10-002, (2010).

\bibitem{CMS-PAS-EGM-10-004}
\href {http://cdsweb.cern.ch/record/1299116} {{ CMS} Collaboration, ``Electron
  Reconstruction and Identification at $\sqrt{s} = 7$ {TeV}'',} CMS Physics
  Analysis Summary CMS-PAS-EGM-10-004, (2010).

\bibitem{antikt}
\hrefCMSnoop {} {M.~Cacciari, G.~P. Salam, and G.~Soyez, ``{The anti-$k_t$ jet
  clustering algorithm}'',} \textit{ JHEP} \textbf{ 04} (2008) 063,
  \href{http://dx.doi.org/10.1088/1126-6708/2008/04/063}{\doi{10.1088/1126-6708/2008/04/063}},
\href{http://www.arXiv.org/abs/0802.1189}{\texttt{ arXiv:0802.1189}}.

\bibitem{JESpaper}
\hrefCMSnoop {} {{ {CMS}} Collaboration, ``Determination of Jet Energy
  Calibration and Transverse Momentum Resolution in {CMS}'',} \textit{ JINST}
  \textbf{ 6} (2011) 11002,
  \href{http://dx.doi.org/10.1088/1748-0221/6/11/P11002}{\doi{10.1088/1748-0221/6/11/P11002}},
  \href{http://www.arXiv.org/abs/1107.4277}{\texttt{ arXiv:1107.4277}}.

\bibitem{CMS-PAS-TOP-10-002}
\hrefCMSnoop {} {{ CMS} Collaboration, ``Measurement of the $\rm t\bar{t}$
  production cross section in pp collisions at 7 TeV using the kinematic
  properties of events with leptons and jets'',} \textit{ Eur. Phys. J. C}
  \textbf{ 9} (2011) 1721,
  \href{http://dx.doi.org/10.1140/epjc/s10052-011-1721-3}{\doi{10.1140/epjc/s10052-011-1721-3}},
  \href{http://www.arXiv.org/abs/1106.0902}{\texttt{ arXiv:1106.0902}}.

\bibitem{CMS-PAS-TOP-10-003}
\hrefCMSnoop {} {{CMS Collaboration}, ``Measurement of the $\rm t\bar{t}$
  production cross section in pp collisions at 7 TeV in lepton+jets events
  using b-quark jet identification'',} \textit{ Phys. Rev. D} \textbf{ 84}
  (2011)
  \href{http://dx.doi.org/10.1103/PhysRevD.84.092004}{\doi{10.1103/PhysRevD.84.092004}},
  \href{http://www.arXiv.org/abs/1108.3773}{\texttt{ arXiv:1108.3773}}.

\bibitem{ThesisPetra}
\href {http://cdsweb.cern.ch/record/1308729} {P.~{Van Mulders}, ``Calibration
  of the jet energy scale using top quark events at the LHC''}.
\newblock PhD thesis, Vrije Universiteit Brussel, 2010.
\newblock \url{http://cdsweb.cern.ch/record/1308729}.

\bibitem{KinFitNote}
\href {http://cdsweb.cern.ch/record/926540} {J.~D'Hondt, S.~Lowette,
  O.~Buchm{\"u}ller{ et~al.}, ``Fitting of Event Topologies with External
  Kinematic Constraints in {CMS}'',} CMS Note 2006/023, (2006).

\bibitem{CMS-PAS-BTV-11-003}
\href {http://cdsweb.cern.ch/record/1421611} {{ CMS} Collaboration,
  ``Measurement of the b-tagging efficiency using \ttbar\ events'',} CMS
  Physics Analysis Summary CMS-PAS-BTV-11-003, (2011).

\bibitem{CMS-PAS-BTV-11-001}
\href {https://cdsweb.cern.ch/record/1366061} {{ {CMS}} Collaboration,
  ``Performance of b-jet identification in {CMS}'',} CMS Physics Analysis
  Summary CMS-PAS-BTV-11-001, (2011).

\bibitem{CMS-PAPERS-TOP-11-002}
\hrefCMSnoop {} {{ {CMS}} Collaboration, ``Measurement of the top-quark
  pair-production cross section and the top-quark mass in the dilepton channel
  at $\sqrt{s} =7$ {TeV}'',} \textit{ JHEP} \textbf{ 07} (2011) 049,
  \href{http://dx.doi.org/10.1007/JHEP07(2011)049}{\doi{10.1007/JHEP07(2011)049}}.

\bibitem{ATLAS-TOPMASS}
\hrefCMSnoop {} {{ ATLAS} Collaboration, ``Measurement of the top quark mass
  with the template method in the top antitop $\rightarrow$ lepton + jets
  channel using ATLAS data'',} (2012).
  \href{http://www.arXiv.org/abs/1203.5755}{\texttt{ arXiv:1203.5755}}.
  Submitted to \textit{Eur. Phys. J. C}.

\bibitem{TopMassCDF}
\hrefCMSnoop {} {{CDF Collaboration}, ``Top Quark Mass Measurement in the
  Lepton + Jets Channel Using a Matrix Element Method and in situ Jet Energy
  Calibration'',} \textit{ Phys. Rev. Lett.} \textbf{ 105} (2010) 252001,
  \href{http://dx.doi.org/10.1103/PhysRevLett.105.252001}{\doi{10.1103/PhysRevLett.105.252001}},
  \href{http://www.arXiv.org/abs/1010.4582}{\texttt{ arXiv:1010.4582}}.

\bibitem{TopMassD0}
\hrefCMSnoop {} {{D0 Collaboration}, ``Precise measurement of the top-quark
  mass from lepton+jets events at D0'',} \textit{ Phys. Rev. D} \textbf{ 84}
  (2011) 032004,
  \href{http://dx.doi.org/10.1103/PhysRevD.84.032004}{\doi{10.1103/PhysRevD.84.032004}},
  \href{http://www.arXiv.org/abs/1105.6287}{\texttt{ arXiv:1105.6287}}.

\bibitem{Jackknife}
\hrefCMSnoop {} {R.~G. Miller, ``The jackknife -- a review'',} \textit{
  Biometrika} \textbf{ 61} (1974) 1.

\bibitem{WZ}
\hrefCMSnoop {} {{CMS Collaboration}, ``Measurements of inclusive W and Z cross
  sections in pp collisions at $\sqrt{s}$ = 7~TeV'',} \textit{ JHEP} \textbf{
  01} (2011) 080,
  \href{http://dx.doi.org/10.1007/JHEP01(2011)080}{\doi{10.1007/JHEP01(2011)080}},
  \href{http://www.arXiv.org/abs/1107.4789}{\texttt{ arXiv:1107.4789}}.

\bibitem{ChargeAsymmetry}
\hrefCMSnoop {} {{CMS Collaboration}, ``Measurement of the lepton charge
  asymmetry in inclusive W production in pp collisions at $\sqrt{s}$ =
  7~TeV'',} \textit{ JHEP} \textbf{ 04} (2011) 050,
  \href{http://dx.doi.org/10.1007/JHEP04(2011)050}{\doi{10.1007/JHEP04(2011)050}},
  \href{http://www.arXiv.org/abs/1103.3470}{\texttt{ arXiv:1103.3470}}.

\bibitem{MuonPerformance}
\hrefCMSnoop {} {{CMS Collaboration}, ``Performance of {CMS} Muon
  Reconstruction in Cosmic-Ray Events'',} \textit{ JINST} \textbf{ 05} (2010)
  T03022,
  \href{http://dx.doi.org/10.1088/1748-0221/5/03/T03022}{\doi{10.1088/1748-0221/5/03/T03022}},
  \href{http://www.arXiv.org/abs/0911.4994}{\texttt{ arXiv:0911.4994}}.

\bibitem{CTEQ}
\hrefCMSnoop {} {P.~M. Nadolsky, H.-L. Lai, Q.-H. Cao{ et~al.}, ``Implications
  of CTEQ global analysis for collider observables'',} \textit{ Phys. Rev. D}
  \textbf{ 78} (2008)
  \href{http://dx.doi.org/10.1103/PhysRevD.78.013004}{\doi{10.1103/PhysRevD.78.013004}}.

\end{thebibliography}\endgroup

\cleardoublepage \appendix\section{The CMS Collaboration \label{app:collab}}\begin{sloppypar}\hyphenpenalty=5000\widowpenalty=500\clubpenalty=5000\textbf{Yerevan Physics Institute,  Yerevan,  Armenia}\\*[0pt]
S.~Chatrchyan, V.~Khachatryan, A.M.~Sirunyan, A.~Tumasyan
\vskip\cmsinstskip
\textbf{Institut f\"{u}r Hochenergiephysik der OeAW,  Wien,  Austria}\\*[0pt]
W.~Adam, T.~Bergauer, M.~Dragicevic, J.~Er\"{o}, C.~Fabjan, M.~Friedl, R.~Fr\"{u}hwirth, V.M.~Ghete, J.~Hammer\cmsAuthorMark{1}, N.~H\"{o}rmann, J.~Hrubec, M.~Jeitler, W.~Kiesenhofer, M.~Krammer, D.~Liko, I.~Mikulec, M.~Pernicka$^{\textrm{\dag}}$, B.~Rahbaran, C.~Rohringer, H.~Rohringer, R.~Sch\"{o}fbeck, J.~Strauss, A.~Taurok, F.~Teischinger, P.~Wagner, W.~Waltenberger, G.~Walzel, E.~Widl, C.-E.~Wulz
% ...
\vskip\cmsinstskip
\dag:~Deceased\\
1:~~Also at CERN, European Organization for Nuclear Research, Geneva, Switzerland\\
% ...

\end{sloppypar}
\end{document}